\documentclass[10pt,aps,prd,showpacs,twocolumn,superscriptaddress,nofootinbib,floatfix]{revtex4-1}
\usepackage{amsfonts,amsmath,amssymb,amsthm}
\usepackage[bookmarks, bookmarksopen, bookmarksnumbered]{hyperref}
\usepackage{graphicx}
\usepackage{tensor}
\usepackage{color}
\usepackage{epigraph}

\begin{document}

\title{Self-gravitating perfect-fluid tori around black holes: Bifurcations, ergoregions, and geometrical properties}

\author{Wojciech Dyba}
\affiliation{Instytut Fizyki im.~Mariana Smoluchowskiego, Uniwersytet Jagiello\'nski, {\L}ojasiewicza 11, 30-348 Krak\'{o}w, Poland}

\author{Wojciech Kulczycki}
\affiliation{Instytut Fizyki im.~Mariana Smoluchowskiego, Uniwersytet Jagiello\'nski, {\L}ojasiewicza 11, 30-348 Krak\'{o}w, Poland}

\author{Patryk Mach}
\affiliation{Instytut Fizyki im.~Mariana Smoluchowskiego, Uniwersytet Jagiello\'nski, {\L}ojasiewicza 11, 30-348 Krak\'{o}w, Poland}

\begin{abstract}
We investigate models of stationary, selfgravitating, perfect-fluid tori (disks) rotating around black holes, focusing on geometric properties of spacetime. The models are constructed within the general-relativistic hydrodynamics, assuming differential (Keplerian) rotation of the fluid. We discuss a parametric bifurcation occurring in the solution space, different possible configurations of ergoregions (including toroidal ergoregions associated with the tori), nonmonotonicity of the circumferential radius, as well as the impact of the torus gravity on the location of the innermost stable circular orbit.  
\end{abstract}

\maketitle

\section{Introduction}

The motivation for studying stationary self-gravitating gaseous tori (disks) around black holes is twofold. On one hand, there is the observational context. Compact tori can form around black holes in a number of astrophysical scenarios, including the gravitational collapse of a supermassive star \cite{shibata2016,sun2017}, or the merger of binary systems consisting of a black hole and a neutron star or two neutron stars \cite{baiotti2017}. The latter are particularly interesting due to the recent detection of gravitational waves attributed to a binary neutron star merger \cite{GW170817}. The tori formed in these events exist for a limited time, which however can be much larger than the dynamical time scale characteristic for such systems.

On the other hand, there is a clear theoretical context. Solutions of the Einstein equations representing self-gravitating tori around black holes provide realistic examples of non-vacuum black-hole spacetimes, which can exhibit interesting properties.

The general relativistic formalism describing stationary perfect-fluid tori around black holes was developed initially by Bardeen already in early 1970's \cite{leshouches}, but the first numerical models were constructed in a series of works by Nishida and Eriguchi \cite{nishida1,nishida2} around 1990's. The numerical scheme used in \cite{nishida1,nishida2} is known as Komatsu-Eriguchi-Hachisu's (KEH) self-consistent field method, as it is based on earlier works dealing with the structure of general-relativistic rotating stars \cite{keh1,keh2}. In the KEH method, the main set of Einstein equations is solved using suitable Green functions. The black hole is introduced assuming suitable boundary conditions. The conditions adopted in \cite{nishida1,nishida2} correspond to a Killing horizon, and they were also derived by Bardeen in \cite{leshouches}.

A highly accurate multi-domain pseudospectral numerical scheme was developed by Ansorg and Petroff in 2005 \cite{ansorg}. It was used later to discuss the possibility that the Komar mass associated with the black hole attains negative values \cite{ansorg:2006}. In these works the implementation of the black hole boundary conditions also followed \cite{leshouches}.

A novel approach, adjusted to the puncture formalism of \cite{brandt:1997}, was presented by Shibata in \cite{shibata}. The black hole horizon was constructed as a Marginally Outer Trapped Surface (MOTS), which in the chosen gauge happened to be a minimal surface. Our numerical setup follows the formalism of \cite{shibata}. It was developed further in three papers co-authored by two of us \cite{kkmmop, kkmmop2, kmm}, where we were mostly focused on the implementation of the Keplerian rotation in self-gravitating configurations. First stationary solutions representing magnetised self-gravitating perfect fluid tori around black holes were constructed in \cite{mgfop}.

Other notable works dealing with stationary, self-gravitating perfect-fluid tori around black holes include \cite{stergioulas} and \cite{tsokaros}.

In this paper, we focus on a collection of interesting features characteristic for sufficiently massive axially symmetric perfect fluid tori rotating around black holes. Since the theory of self-gravitating, general-relativistic perfect-fluid tori around black holes is almost intractable analytically, all these features are demonstrated with specific numerical solutions. We assume the Keplerian rotation law introduced in \cite{kkmmop, kkmmop2} and, for simplicity, the polytropic equation of state.

We start by discussing a bifurcation in the solution space. We fix black-hole parameters, the polytropic exponent, inner and outer coordinate equatorial radii of the torus, as well as its maximal rest-mass density and find two solutions differing in the asymptotic mass. There is a branch of solutions corresponding to relatively light tori and another one with the mass of the tori exceeding many times the mass of the central black hole.

The next effect has already been discovered by Labranche, Petroff, and Ansorg for rigidly rotating strange matter tori (with an equation of state based on the MIT bag model) and without the central object \cite{labranche}. It turns out that for sufficiently massive tori, the relation between the coordinate radius and the circumferential radius at the equatorial plane is not monotonic. There is a local maximum of the circumferential radius, which occurs inside the torus. Thus, the largest circle (in the sense of its circumference) that can be embedded within the torus is not the outermost one, but one located inside the torus. This also suggests that the proper volume of the torus could be much bigger than an estimate based on the values of the torus inner and outer radii. This expectation is confirmed with numerical examples. Because of the axially symmetric context we refer to the above effect as the breaking of the Pappus-Guldinus rule.

Subsequently, we discuss the ergospheres (or rather ergoregions) in the case with massive tori. It was known at least since the analysis of \cite{neugebauer,meinel:1995} that a massive, rotating disk can have its own ergoregion. In \cite{ansorg:2006} Petoroff and Ansorg showed numerical examples of rigidly rotating perfect fluid tori around black holes exhibiting two disconnected ergoregions: one with the spherical topology, connected with the black hole, and a toroidal one, associated with the torus. There is also a parametric transition to the case with a single ergoregion of spherical topology, encompassing both the black hole and the torus. In this paper we give examples of similar behavior for more realistic, Keplerian rotation law and polytropic equations of state.

The next effect is of physical interest for moderately heavy toroids. We investigate the location of the Innermost Stable Circular Orbit (ISCO) of a test particle in the obtained spacetimes. It turns out to be quite sensitive to the self-gravity of the torus---the discrepancy between the true location of the ISCO, and the corresponding location in the Kerr spacetime can be nonnegligible, even for relatively light toroids. A sufficiently massive torus can create an additional region, outside the ISCO, in which the circular geodesics are unstable. For even heavier tori, there can occur regions in which no circular geodesic orbits can exist. Investigating the geodesic motion around distorted black holes is a natural and active field of research, and focusing on the location of the ISCO is motivated by the astrophysical context. An example of a quite general, recent analysis in this direction is given in \cite{shoom}.

The order of this paper is as follows. The next section contains a description of the model. We recall the relevant formulas from \cite{shibata,kkmmop,kkmmop2}. The parametric bifurcation is discussed in Sec.\ \ref{secbif}. In Section \ref{pappus_guldinus} we give examples of nonmonotonicity of the circumferential radius and discuss the breaking of the Pappus-Guldinus rule. Examples of ergoregions occurring for solutions with sufficiently massive tori are given in Sec.\ \ref{secergo}. In Section \ref{circgeodesics} we analyze equatorial, circular geodesics in the obtained spacetimes, focusing on the location of the ISCO. Finally, concluding remarks are given in Sec.\ \ref{conclusions}. 

\section{Stationary toroids around black holes}

The formulation used in this paper is based on  a scheme developed in \cite{shibata}, and used (with some modifications) also in \cite{kkmmop,kkmmop2,kmm}. In this section we only give a brief description of the key elements of the formalism.

Except for few places, we use standard gravitational system of units with $c = G = 1$, where $c$ is the speed of light, and $G$ is the gravitational constant. The signature of the metric is assumed to be $(-,+,+,+)$. In what follows Greek indices are used to label spacetime dimensions $\mu = 0,1,2,3$. Spatial dimensions are labeled with Latin indices $i = 1,2,3$.

\subsection{Euler-Bernoulli equation}

We will work in spherical coordinates $(t, r , \theta, \varphi)$. We assume a stationary, axially symmetric metric of the form
\begin{equation}
\label{generalmetric}
g = g_{tt} dt^2 + 2 g_{t \varphi} dt d\varphi + g_{rr} dr^2 + g_{\theta \theta} d\theta^2 + g_{\varphi \varphi} d\varphi^2, 
\end{equation}
where functions $g_{tt}$, $g_{t\varphi}$, $g_{rr}$, $g_{\theta \theta}$, $g_{\varphi \varphi}$ depend only on $r$ and $\theta$. When necessary, we will also specialize to the following quasi-isotropic form
\begin{eqnarray}
\nonumber
g & = & - \alpha^2 dt^2 + \psi^4 e^{2q} (dr^2 + r^2 d\theta^2) + \\
&& \psi^4 r^2 \sin^2 \theta (\beta dt + d \varphi)^2.
\label{isotropic}
\end{eqnarray}
Both forms of the metric admit two independent Killing vectors, azimuthal (rotational) and asymptotically timelike, with contravariant components $\eta^\mu = (0,0,0,1)$ and $\xi^\mu = (1,0,0,0)$, respectively.

We assume the energy-momentum tensor of the perfect fluid
\begin{equation}
T^{\mu \nu} = \rho h u^\mu u^\nu + p g^{\mu \nu},
\end{equation}
where $\rho$ is the rest-mass density, $h$ is the specific enthalpy, $p$ is the thermal pressure, $u^\mu$ denotes the four-velocity of the fluid, and $g_{\mu \nu}$ is the metric tensor. The four-velocity $u^\mu$ is normalized: $u_\mu u^\mu = -1$.

In the following, we only consider the simplest stationary rotation with the four-velocity in the form $u^\mu = (u^t,0,0,u^\varphi) = u^t(1,0,0,\Omega)$. The component $u^t$ can be expressed in terms of the angular velocity $\Omega = u^\varphi/u^t$ as
\begin{equation}
\label{ut}
(u^t)^2 = - \frac{1}{g_{tt} + 2 g_{t \varphi} \Omega + g_{\varphi \varphi} \Omega^2}.
\end{equation}
Because of the assumptions of stationarity and axial symmetry, $u^t$, $u^\varphi$, $\rho$, $p$, and $h$ can only depend on $r$ and $\theta$.

For a barotropic fluid, the conservation equations
\begin{equation}
\label{cons_eqs_general}
\nabla_\mu \left( \rho u^\mu \right) = 0, \quad \nabla_\mu T^{\mu \nu} = 0
\end{equation}
can be integrated, provided that the angular momentum per unit inertial mass, $j = u^t u_\varphi$, is a function of the angular velocity $\Omega$ only. In this case, one gets (in the region where $\rho > 0$)
\begin{equation}
\label{bernoulli3}
\ln \left( \frac{h}{u^t} \right) + \int j(\Omega) d\Omega  = C,
\end{equation}
where $C$ denotes an integration constant. We will refer to Eq.\ (\ref{bernoulli3}) as the Euler-Bernoulli equation.

The solutions of this paper are computed assuming the Keplerian rotation law 
\begin{eqnarray}
j(\Omega) & = & \frac{a^2 \Omega^\frac{4}{3} + w^\frac{4}{3} (1 - 3 a \Omega) (1 - a \Omega)^\frac{1}{3}}{\Omega^\frac{1}{3} \left[ 1 - a^2 \Omega^2 - 3 w^\frac{4}{3} \Omega^\frac{2}{3}
   (1-a \Omega)^\frac{4}{3} \right]} \nonumber \\
& = & - \frac{1}{2} \frac{d}{d \Omega} \ln \left\{ 1 - \left[ a^2 \Omega^2 + 3 w^\frac{4}{3} \Omega^\frac{2}{3} (1 - a \Omega)^\frac{4}{3} \right] \right\}.
\label{keplerian_rl}
\end{eqnarray}
The derivation of this formula and the discussion of its physical relevance was given in \cite{kkmmop, kkmmop2}. This rotation law was also used in \cite{kmm, mgfop}. Here $w$ is a free constant, and $a$ is a spin parameter of the black hole defined in Sec.\ \ref{einstein_equations}. The circular geodesic motion in the equatorial plane of the Kerr spacetime with the mass $m$ and spin $a$ is characterized by Eq.\ (\ref{keplerian_rl}) with $w^2 = m$. However in general, $w^2 \neq m$. The value of $w$ is obtained by demanding that the inner and outer equatorial coordinate radii of the disks are given by $r_1$ and $r_2$, respectively. Note that in the Newtonian limit, Eq.\ (\ref{keplerian_rl}) yields $\Omega = w/(r \sin \theta)^\frac{3}{2}$.

Given the relation $j(\Omega)$ and the metric, one can compute the angular velocity $\Omega$ by solving the equation
\begin{equation}
\label{rot_law_eq}
j(\Omega) \left[ \alpha^2 - \psi^4 r^2 \sin^2 \theta (\Omega + \beta)^2 \right] = \psi^4 r^2 \sin^2 \theta (\Omega + \beta),
\end{equation}
which is a direct consequence of the definition $j = u^t u_\varphi$. In the following, we assume a convention with $\Omega > 0$. We say that the torus is corotating, if $a>0$, and counterrotating, for $a < 0$.

Assuming the above choices, one can write the Euler-Bernoulli Eq.\ (\ref{bernoulli3}) as
\begin{eqnarray}
\label{bernoulli2}
\lefteqn{h \sqrt{ \alpha^2 - \psi^4 r^2 \sin^2 \theta (\Omega + \beta)^2}} \nonumber \\
&& \times \left\{ 1 - \left[ a^2 \Omega^2 + 3 w^\frac{4}{3} \Omega^\frac{2}{3} (1 - a \Omega)^\frac{4}{3} \right] \right\}^{-\frac{1}{2}} = C^\prime,
\end{eqnarray}
where $C^\prime$ is a constant.

We assume a polytropic equation of state $p = K\rho^\Gamma$, where $K$ and $\Gamma$ are constant. The specific enthalpy is then given by
\begin{eqnarray}
\label{polytropic_h}
h = 1 + \frac{K \Gamma}{\Gamma - 1} \rho^{\Gamma - 1}.
\end{eqnarray}

\subsection{Einstein equations}
\label{einstein_equations}

Although for self-gravitating tori the spacetime metric is not given by the Kerr solution, the Kerr metric plays an important role in the following construction. For completeness, we write the Kerr metric in the quasi-isotropic coordinates of the form (\ref{isotropic}) \cite{shibata,brandtseidel}. Define
\begin{eqnarray}
r_\mathrm{K} & = & r \left( 1 + \frac{m}{r} + \frac{m^2 - a^2}{4 r^2} \right), \\ 
\Delta_\mathrm{K} & = & r_\mathrm{K}^2 -2r_\mathrm{K}+a^2, \\
\Sigma_\mathrm{K} & = & r_\mathrm{K}^2 + a^2 \cos^2 \theta,
\end{eqnarray}
where $m$ and $a m$ denote the asymptotic mass and angular momentum of the Kerr spacetime, respectively. The Kerr metric can be expressed as
\begin{eqnarray}
g & = & - \alpha_\mathrm{K}^2 dt^2 + \psi_\mathrm{K}^4 e^{2q_\mathrm{K}} (dr^2 + r^2 d\theta^2) + \nonumber \\
&& \psi_\mathrm{K}^4 r^2 \sin^2 \theta (\beta_\mathrm{K} dt + d \varphi)^2,
\end{eqnarray}
where
\begin{eqnarray}
\psi_\mathrm{K} & = & \frac{1}{\sqrt{r}}\Bigl( r^2_\mathrm{K}  +a^2 +2ma^2\frac{r_\mathrm{K}\sin^2\theta  }{\Sigma_\mathrm{K}}\Bigr)^{1/4}, \\
\beta_\mathrm{K} & = & -\frac{2mar_\mathrm{K}}{(r^2_\mathrm{K}+a^2)\Sigma_\mathrm{K} +2ma^2r_\mathrm{K} \sin^2\theta}, \\
\alpha_\mathrm{K} & = & \left[ \frac{ \Sigma_\mathrm{K} \Delta_\mathrm{K}}{(r_\mathrm{K}^2+a^2)\Sigma_\mathrm{K}+2ma^2r_\mathrm{K} \sin^2\theta} \right]^{1/2}, \label{eq:alpha_K}\\
e^{q_\mathrm{K}} & = & \frac{\Sigma_\mathrm{K}}{\sqrt{(r^2_\mathrm{K}+a^2)\Sigma_\mathrm{K} +2ma^2r_\mathrm{K} \sin^2\theta}}.
\end{eqnarray}

In the following, we adopt the puncture formalism, as described by \cite{shibata}. Let $m$ and $a$ be parameters, corresponding to some Kerr spacetime. We define $r_\mathrm{s} = \frac{1}{2}\sqrt{m^2 - a^2}$, so that for the Kerr metric with the asymptotic mass $m$ and the asymptotic angular momentum $am$, the event horizon would be a sphere $r = r_\mathrm{s}$.  Returning to the general, self-gravitating case, we replace the functions $\psi$ and $\alpha$ (the lapse) by $\phi$ and $B$ defined by the following relations
\begin{equation}
\label{puncture}
\psi = \left( 1 + \frac{r_\mathrm{s}}{r} \right) e^\phi, \quad \alpha \psi = \left( 1 - \frac{r_\mathrm{s}}{r} \right) e^{-\phi} B.
\end{equation}

The shift vector is split in two parts, $\beta = \beta_\mathrm{K} + \beta_\mathrm{T}$, as follows. We write the only non-vanishing components of the extrinsic curvature of the slices of constant time $t$ as
\begin{equation}
\label{Krf}
K_{r \varphi} = K_{\varphi r} = \frac{H_\mathrm{E} \sin^2 \theta}{\psi^2 r^2} +  \frac{1}{2 \alpha} \psi^4 r^2 \sin^2 \theta \partial_r \beta_\mathrm{T},
\end{equation}
\begin{equation}
\label{Ktf}
K_{\theta \varphi} = K_{\varphi \theta} = \frac{H_\mathrm{F} \sin \theta}{\psi^2 r} + \frac{1}{2 \alpha} \psi^4 r^2 \sin^2 \theta \partial_\theta \beta_\mathrm{T},
\end{equation}
where $H_\mathrm{E}$ and $H_\mathrm{F}$ are given by
\begin{eqnarray}
H_\mathrm{E} & = & \frac{ma \left[ (r_\mathrm{K}^2 - a^2) \Sigma_\mathrm{K} + 2 r_\mathrm{K}^2 (r_\mathrm{K}^2 + a^2) \right]}{\Sigma_\mathrm{K}^2}, \\
H_\mathrm{F} & = & - \frac{2 m a^3 r_\mathrm{K} \sqrt{\Delta_\mathrm{K}} \cos \theta \sin^2 \theta}{\Sigma_\mathrm{K}^2}.
\end{eqnarray}
Equations (\ref{Krf}) and (\ref{Ktf}) can serve as an implicit definition of $\beta_\mathrm{T}$. It can be checked that for the Kerr spacetime, $\beta_\mathrm{T} = 0$. In a sense, $\beta_\mathrm{K}$ is associated with the black hole, while $\beta_\mathrm{T}$ corresponds to the torus.

The Einstein equations can be written as the following system of equations for the functions $q$, $\phi$, $B$ and $\beta_\mathrm{T}$:
\begin{widetext}
\begin{subequations}
\label{main_sys}
\begin{eqnarray}
\left[ \partial_{rr} + \frac{1}{r } \partial_r  + \frac{1}{r^2} \partial_{\theta \theta}  \right] q & = & S_q, \label{47}\\
\left[ \partial_{rr} + \frac{2 r  }{r^2 - r_\mathrm{s}^2} \partial_r + \frac{1}{r^2} \partial_{\theta \theta} + \frac{  \cot{\theta}}{r^2}  \partial_\theta \right] \phi & = & S_\phi, \label{44} \\
\left[ \partial_{rr} + \frac{3 r^2 +  r_\mathrm{s}^2}{r(r^2 - r_\mathrm{s}^2)} \partial_r + \frac{1}{r^2} \partial_{\theta \theta} + \frac{2 \cot{\theta}}{r^2}  \partial_\theta \right] B & = & S_B, \label{45} \\
\left[ \partial_{rr} + \frac{4 r^2 - 8 r_\mathrm{s} r + 2 r_\mathrm{s}^2}{r(r^2 - r_\mathrm{s}^2)} \partial_r + \frac{1}{r^2} \partial_{\theta \theta} + \frac{3 \cot{\theta}}{r^2}  \partial_\theta \right]  \beta_\mathrm{T} & = & S_{\beta_\mathrm{T}},  \label{46}
\end{eqnarray}
 \end{subequations}
where
\begin{subequations}
\label{sources}
\begin{eqnarray}
S_q & = & -8 \pi e^{2q} \left( \psi^4 p - \frac{\rho h u_\varphi^2}{r^2 \sin^2 \theta} \right) + \frac{3 A^2}{\psi^8} + 2 \left[ \frac{r - r_\mathrm{s}}{r(r + r_\mathrm{s})} \partial_r + \frac{\cot \theta}{r^2} \partial_\theta \right] b + \left[ \frac{8 r_\mathrm{s}}{r^2 - r_\mathrm{s}^2} + 4 \partial_r (b - \phi) \right] \partial_r \phi \nonumber \\
& & + \frac{4}{r^2} \partial_\theta \phi \partial_\theta (b - \phi), \\
S_\phi & = & - 2 \pi e^{2q} \psi^4 \left( \rho_\mathrm{H} - p + \frac{\rho h u_\varphi^2}{\psi^4 r^2 \sin^2 \theta} \right) - \frac{A^2}{\psi^8} - \partial_r\phi \partial_r b - \frac{1}{r^2} \partial_\theta \phi \partial_\theta b - \frac{1}{2} \left[ \frac{r - r_\mathrm{s}}{r (r + r_\mathrm{s})} \partial_r b + \frac{\cot \theta}{r^2} \partial_\theta b \right], \\
S_B & = & 16 \pi B e^{2q} \psi^4 p, \\
S_{\beta_\mathrm{T}} & = & \frac{16 \pi \alpha e^{2q} j_\varphi}{r^2 \sin^2 \theta} - 8 \partial_r \phi \partial_r \beta_\mathrm{T} + \partial_r b \partial_r \beta_\mathrm{T} - 8 \frac{\partial_\theta \phi \partial_\theta \beta_\mathrm{T}}{r^2} + \frac{\partial_\theta b \partial_\theta \beta_\mathrm{T}}{r^2},
\end{eqnarray}
\end{subequations}
\end{widetext}
The equation for $\beta_\mathrm{K}$ reads:
\begin{equation}
\label{betak_eq}
\partial_r \beta_\mathrm{K} = 2 H_\mathrm{E} B e^{-8 \phi} \frac{(r - r_\mathrm{s})r^2}{(r + r_\mathrm{s})^7}.
\end{equation}
In the above formulas $B = e^b$ and
\begin{equation}
\label{a2formula}
A^2 = \frac{(\psi^2 K_{r \varphi})^2}{r^2 \sin^2 \theta} + \frac{(\psi^2 K_{\theta \varphi})^2}{r^4 \sin^2 \theta}.
\end{equation}
In addition
\begin{equation}
\rho_\mathrm{H} = \alpha^2 \rho h (u^t)^2 - p,
\end{equation}
\begin{equation}
j_\varphi = \alpha \rho h u^t u_\varphi.
\end{equation}

The boundary conditions at $r = r_\mathrm{s}$ read
\begin{equation}
\partial_r q = \partial_r \phi = \partial_r B = \partial_r \beta_\mathrm{T} = 0.
\end{equation}
Equation (\ref{46}) admits a more stringent boundary condition. Following \cite{shibata} we set $\beta_\mathrm{T} = O[(r - r_\mathrm{s})^4]$, which is equivalent to $\beta_\mathrm{T} = \partial_r \beta_\mathrm{T} = \partial_{rr} \beta_\mathrm{T} = \partial_{rrr} \beta_\mathrm{T} = 0$ at $r = r_\mathrm{s}$.

It can be easily shown that the above conditions guarantee that the two-surface $r = r_\mathrm{s}$ embedded in a hypersurface of constant time $\Sigma_t$ is a minimal surface. In the assumed quasi-isotropic gauge (\ref{isotropic}) it is also a Marginally Outer Trapped Surface (MOTS) or the so-called apparent horizon \cite{mgfop}.

\subsection{Masses and angular momenta}

It is both customary and natural to characterize black hole--torus systems by the masses and angular momenta of their constituents. In this context, the most natural mass measure is the Arnowitt-Deser-Misner (ADM) asymptotic mass. For practical reasons, it is not computed in terms of an asymptotic surface integral. Instead, we compute the ADM mass as
\begin{eqnarray} 
M_\mathrm{ADM} = \sqrt{m^2 - a^2} + M_1, 
\end{eqnarray}
where
\begin{eqnarray} 
M_1 = - 2 \int_{r_\mathrm{s}}^\infty dr \int_0^{\pi/2} d \theta (r^2 - r_\mathrm{s}^2) \sin \theta S_\phi, 
\end{eqnarray}
and $m$ is the black-hole mass parameter introduced in Sec.\ \ref{einstein_equations}.

The mass of the black hole is given by Christodoulou's formula \cite{christodoulou}
\begin{eqnarray}
 M_\mathrm{BH} = M_\mathrm{irr} \sqrt{1 + \frac{J_\mathrm{H}^2}{4 M_\mathrm{irr}^4}}.
\end{eqnarray}
Here $J_\mathrm{H}$ is the angular momentum of the black hole,
\begin{eqnarray}
J_\mathrm{H} = \frac{1}{4} \int_0^{\pi/2} d \theta \left( \frac{r^4 \sin^3 \theta \psi^6 \partial_r \beta}{\alpha} \right)_{r=r_\mathrm{s}}, 
\end{eqnarray}
and $M_\mathrm{irr}$ denotes the so-called irreducible mass, defined as
\begin{eqnarray}
M_\mathrm{irr} = \sqrt{\frac{A_\mathrm{H}}{16 \pi}},
\end{eqnarray}
where $A_\mathrm{H}$ is the area of the horizon,
\begin{eqnarray}
A_\mathrm{H} = 4 \pi \int_0^{\pi/2} d \theta \left( \psi^4 e^q r^2 \sin \theta \right)_{r=r_\mathrm{s}}.
\end{eqnarray}

The angular momentum of the torus is defined as
\begin{eqnarray}
J_1 & = & \int \sqrt{- g} T\indices{^t_\varphi} d^3 x \nonumber \\
& = & 4 \pi \int_{r_\mathrm{s}}^\infty dr \int_0^{\pi/2} d\theta r^2 \sin \theta \alpha \psi^6 e^{2q} \rho h u^t u_\varphi.
\end{eqnarray}
The above definition corresponds to the Killing vector $\eta^\mu = (0,0,0,1)$ and the conservation law $\eta^\nu \nabla_\mu T\indices{^\mu_\nu} = \nabla_\mu (T\indices{^\mu_\nu} \eta^\nu) = 0$ \cite{leshouches}. The total angular momentum can be expressed as
\begin{equation}
J = J_\mathrm{H} + J_1.
\end{equation}

It should be stressed that the value assigned to $J_\mathrm{H}$ depends on the assumed boundary conditions at $r = r_\mathrm{s}$. In our case (and in \cite{shibata}) the condition $\partial_r \beta_\mathrm{T} = 0$ at $r = r_\mathrm{s}$ implies $J_\mathrm{H} = a m$. Note that a  natural definition of the black-hole spin would be
\begin{equation}
\hat a = \frac{J_\mathrm{H}}{M_\mathrm{BH}} = \frac{a m}{M_\mathrm{BH}}.
\end{equation}
In general $\hat a \neq a$.

The discussion of other mass measures (and the relations between them) can be found in \cite{shibata,mgfop}.

\subsection{Parametrization of solutions}

The whole black-hole torus system is described by Eqs.\ (\ref{keplerian_rl}), (\ref{rot_law_eq}), (\ref{bernoulli2}), (\ref{polytropic_h}), (\ref{main_sys}), (\ref{betak_eq}). There are a few ways of parametrizing the solutions of these equations. Here we follow the choice of \cite{kkmmop,kkmmop2,mgfop}. We specify the black hole parameters $m$ and $a$, the inner and outer equatorial coordinate radii of the torus $r_1$ and $r_2$, the polytropic exponent $\Gamma$, and the maximal rest-mass density within the torus $\rho_\mathrm{max}$. This means, in particular, that the constants $w$ in Eq.\ (\ref{keplerian_rl}), $K$ in Eq.\ (\ref{polytropic_h}) and $C^\prime$ in Eq.\ (\ref{bernoulli2}) are computed together with the solution.

\subsection{Numerical method}

The numerical method used in this paper has been described in detail and tested in \cite{kkmmop2}. The only change with respect to the version described in \cite{kkmmop2} is a replacement of linear algebra routines. The previous version of the code used LAPACK \cite{lapack}, which is now replaced by the PARDISO library \cite{pardiso}.

In the remaining sections we discuss properties of the obtained numerical solutions.

\section{Bifurcations}
\label{secbif}

\begin{figure}
\includegraphics[width=\columnwidth]{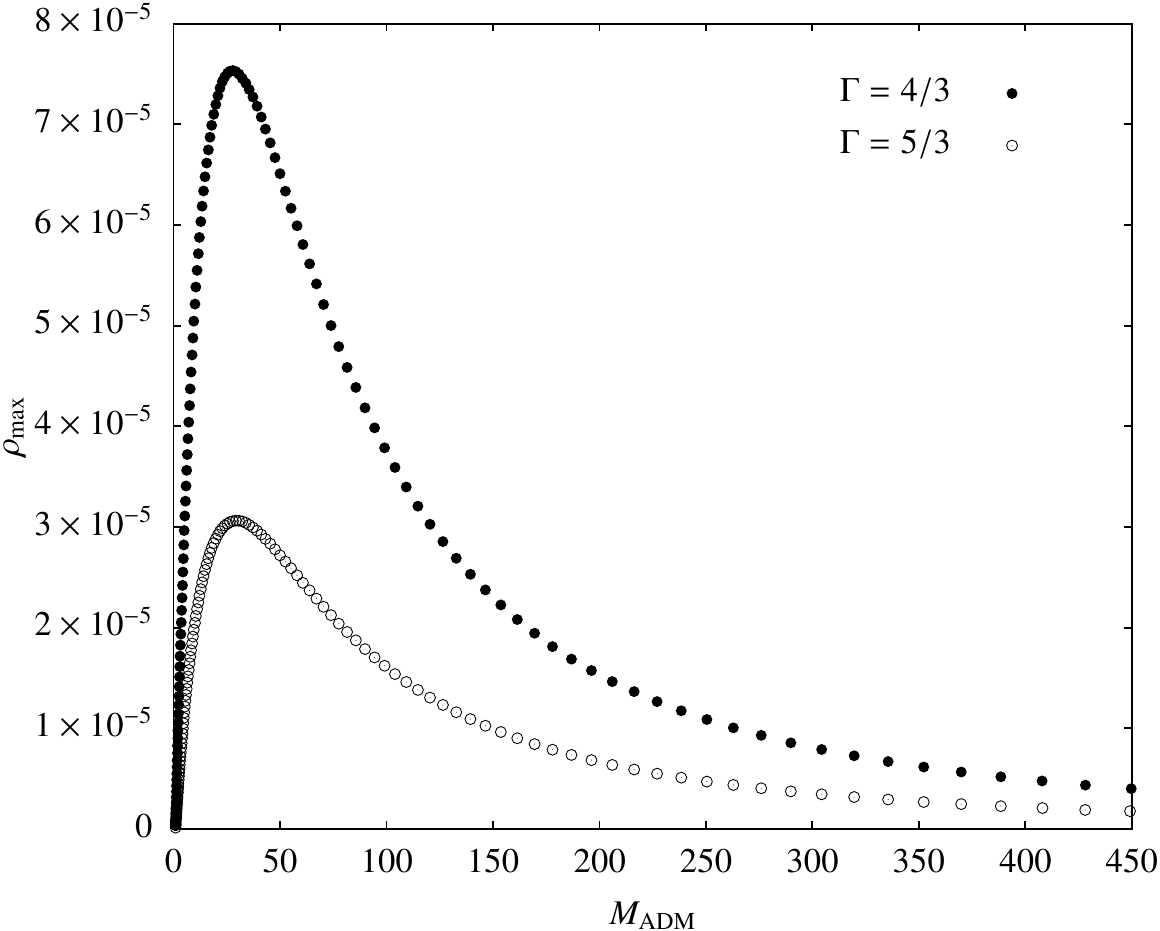}
\caption{\label{biff_r2_100_g43}The maximal density $\rho_\textrm{max}$ versus the ADM mass $M_\textrm{ADM}$. Geometric parameters of the solutions are: $r_1 = 50$, $r_2=100$, $a=0$, $m=1$. Solid dots correspond to the polytropic equation of state with $\Gamma=4/3$; empty circles depict  solutions with $\Gamma=5/3$.}
\end{figure}

\begin{figure}
\includegraphics[width=\columnwidth]{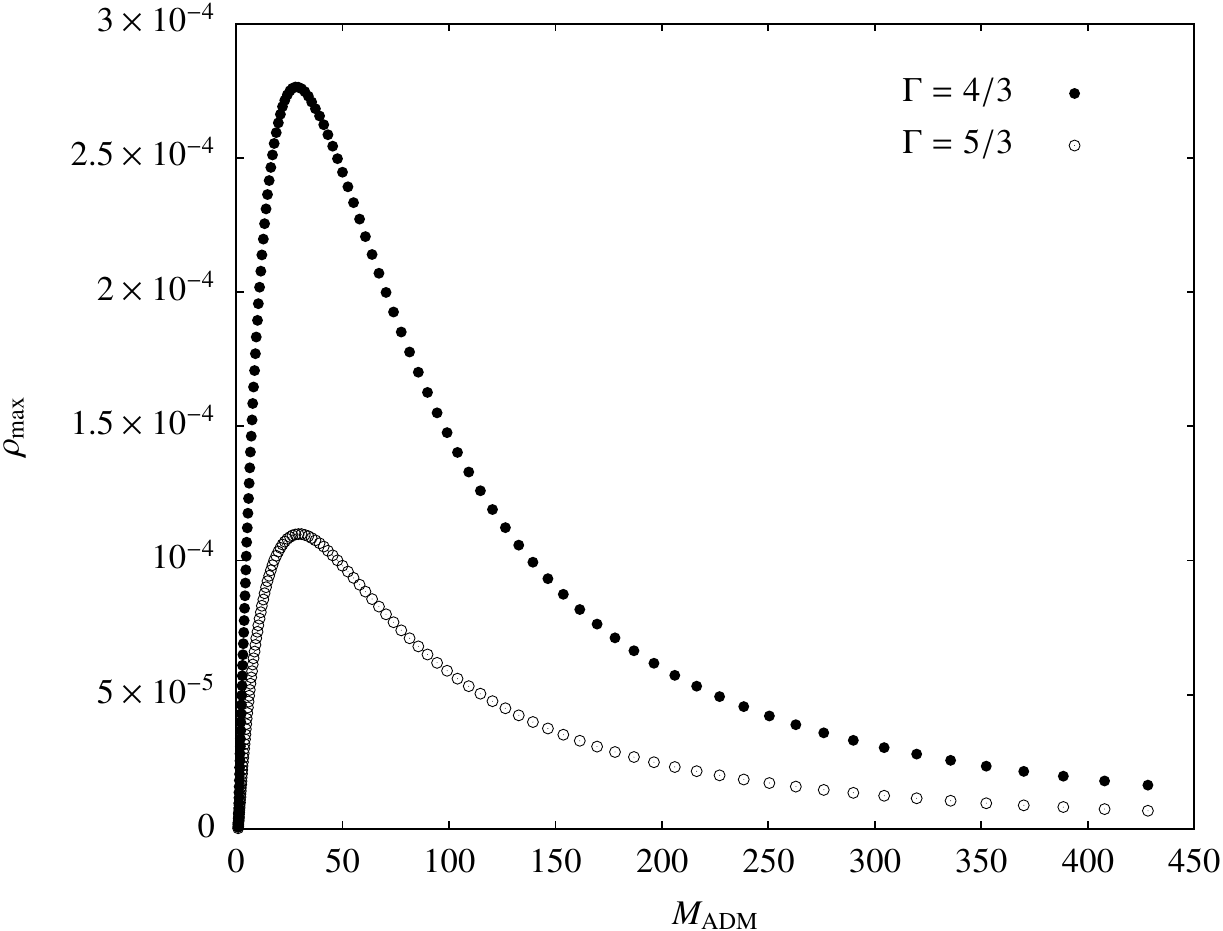}
\caption{\label{biff_r2_10^15_g43}The same as in Fig.\ \ref{biff_r2_100_g43}. Geometric parameters of the solutions are: $r_1 = 75$, $r_2=100$, $a=0$, $m=1$. Solid dots correspond to the polytropic equation of state with $\Gamma=4/3$; empty circles depict solutions with $\Gamma=5/3$.}
\end{figure}

\begin{figure}
\includegraphics[width=\columnwidth]{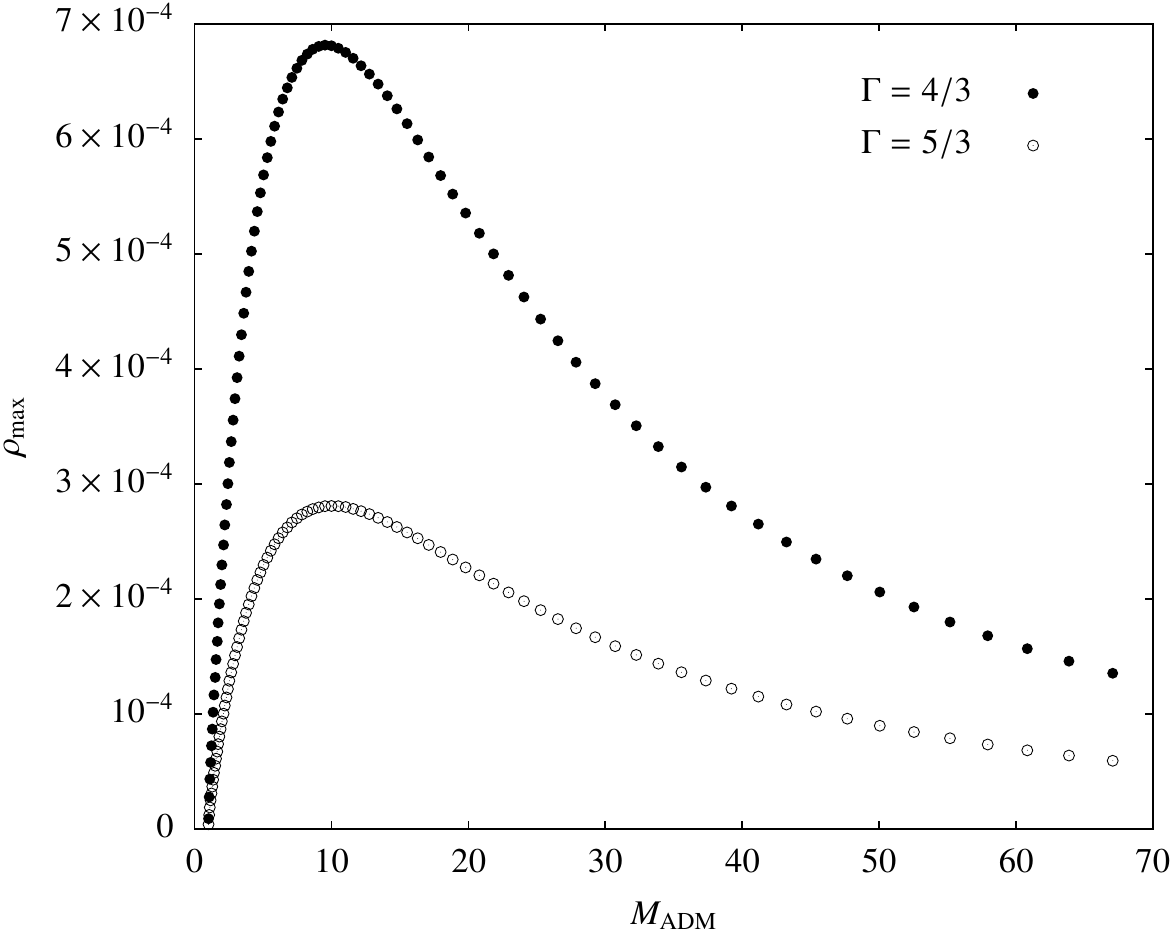}
\caption{\label{biff_r2_100_g53}The same as in Fig.\ \ref{biff_r2_100_g43}. Geometric parameters of the solutions are: $r_1 \approx 15.8$, $r_2 \approx 31.6$, $a=0$, $m=1$. Solid dots correspond to the polytropic equation of state with $\Gamma=4/3$; empty circles depict solutions with $\Gamma=5/3$.}
\end{figure}

\begin{figure}
\includegraphics[width=\columnwidth]{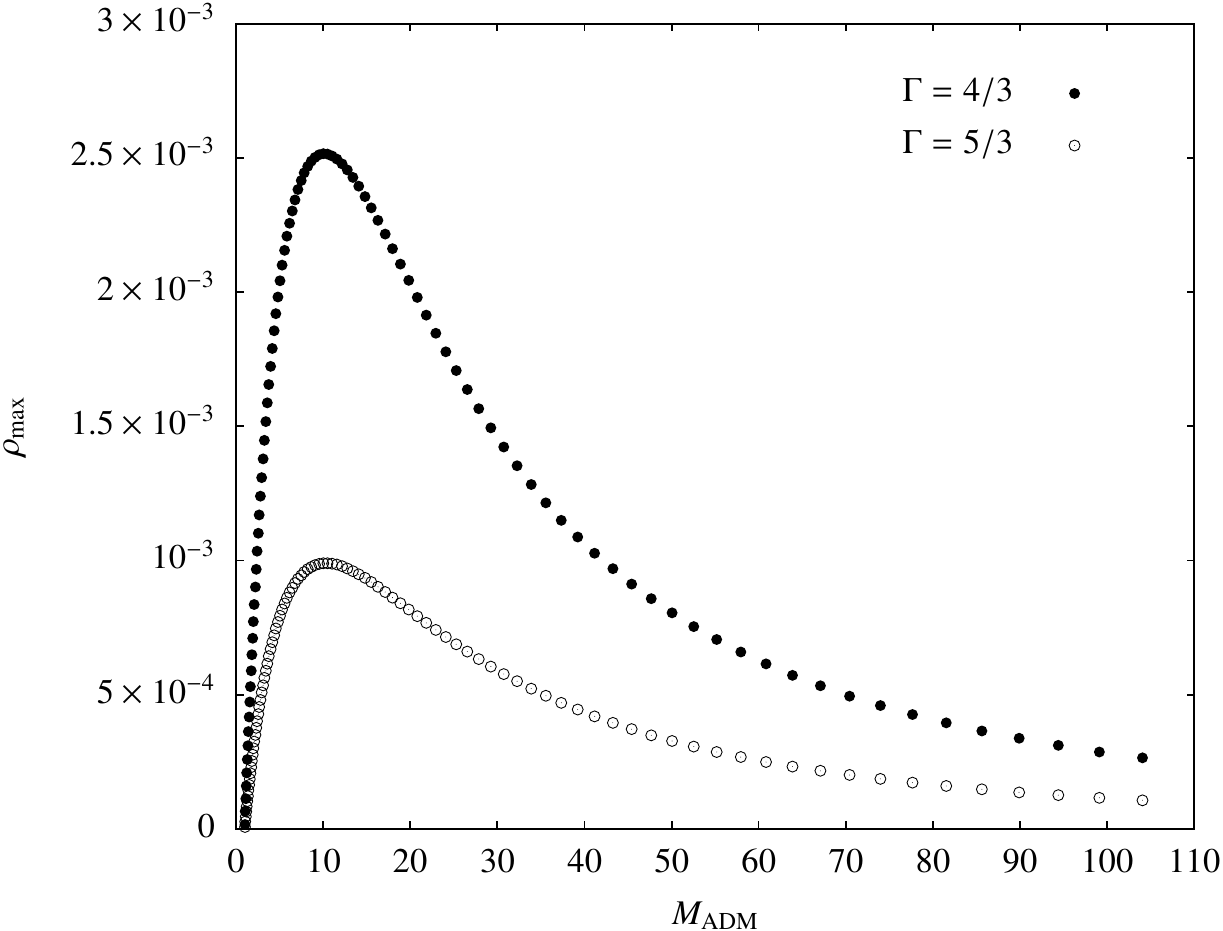}
\caption{\label{biff_r2_10^15_g53} The same as in Fig.\ \ref{biff_r2_100_g43}. Geometric parameters of the solutions are: $r_1 \approx 23.7$, $r_2 \approx 31.6$, $a=0$, $m=1$. Solid dots correspond to the polytropic equation of state with $\Gamma=4/3$; empty circles depict solutions with $\Gamma=5/3$.}
\end{figure}

\begin{figure}
\includegraphics[width=\columnwidth]{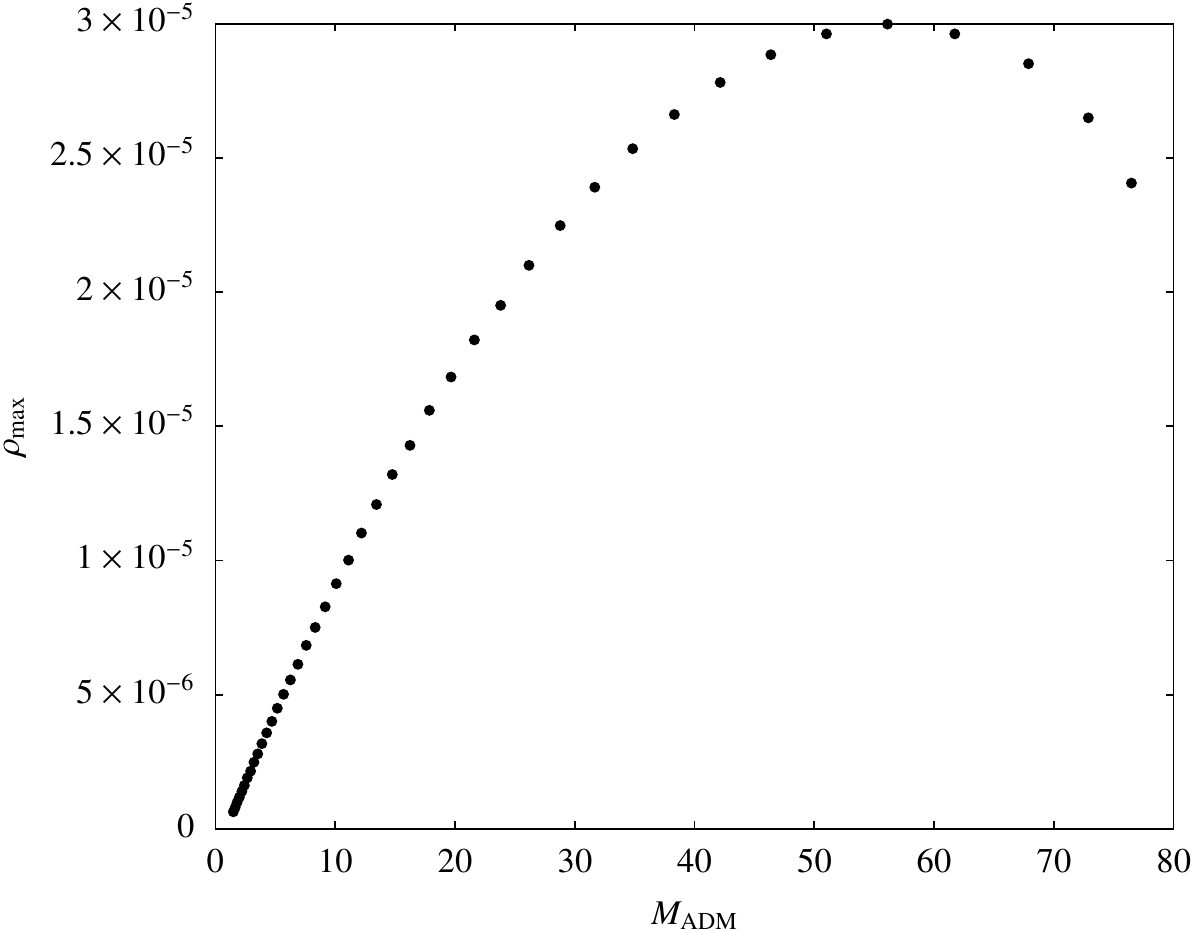}
\caption{\label{biff_rc075_m0_005} The same as in Fig.\ \ref{biff_r2_100_g43}. Geometric parameters of the solutions are: $r_\textrm{c,1} \approx 100$, $r_\textrm{c,2} \approx 200$, $a=0$, $m=1$, $\Gamma=4/3$. Note that instead of fixing coordinate radii $r_1$ and $r_2$, we fix their circumferential values $r_\textrm{c,1}$ and $r_\textrm{c,2}$.}
\end{figure}

The first surprising fact concerning the regime of massive tori is the parametric bifurcation occurring in the space of solutions.

It is convenient to parametrize the solutions by specifying the mass and spin parameters of the black hole $m$ and $a$, the inner and outer equatorial coordinate radii of the torus $r_1$ and $r_2$, the polytropic exponent $\Gamma$, and the maximum density within the torus $\rho_\mathrm{max}$. One would expect to control the total asymptotic mass of the system $M_\mathrm{ADM}$ by changing the value of $\rho_\mathrm{max}$. It turns out that the dependence of $\rho_\mathrm{max}$ on $M_\mathrm{ADM}$ is not monotonic. For light tori, $\rho_\mathrm{max}$ grows with $M_\mathrm{ADM}$; it attains a maximum for a sufficiently large $M_\mathrm{ADM}$, and then decreases with $M_\mathrm{ADM}$ growing further. The latter is of course highly counterintuitive. We illustrate this behavior in Figs.\ \ref{biff_r2_100_g43}--\ref{biff_r2_10^15_g53} for a sample of solutions. The parameters of these solutions are reported in the captions of the figures. Each point in the graphs shown in Figs.\ \ref{biff_r2_100_g43}--\ref{biff_r2_10^15_g53} corresponds to a numerical solution. Solutions with the same parameters $m$, $a$, $\Gamma$, $r_1$, $r_2$ are denoted with the same shape of points.

The first obvious consequence is that for fixed parameters $m$, $a$, $r_1$, $r_2$, and $\Gamma$, there is a limit on the maximal allowed density $\rho_\mathrm{max}$, for which stationary solutions can be found.

This behavior can be partially explained by the fact that $r_1$ and $r_2$ are coordinate radii, devoid of a clear geometrical meaning. In axial symmetry, the most natural geometric measure of the distance from the symmetry axis is probably the so-called circumferential radius---a circle with a circumference $L$ has a circumferential radius $r_\mathrm{c} = L/(2\pi)$. The circumferential radius of a circle of constant $t$, $r$ and $\theta$ in a spacetime with the metric (\ref{generalmetric}) or (\ref{isotropic}) is $r_\mathrm{c} = \sqrt{g_{\varphi \varphi}} = \psi^2 r \sin \theta$. Also note that $r_\mathrm{c}$ can be expressed covariantly in terms of the Killing vector $\eta^\mu$ as $r_\mathrm{c} = \sqrt{\eta_\mu \eta^\mu}$.

One can check, for the data depicted in Figs.\ \ref{biff_r2_100_g43}--\ref{biff_r2_10^15_g53}, that the circumferential radius corresponding to $r_2$ grows with increasing $M_\mathrm{ADM}$ (cf.\ Sec.\ \ref{pappus_guldinus}). More importantly, the proper volume of the torus can also grow with $M_\mathrm{ADM}$. In other words, it is possible to obtain a sequence of tori with a decreasing rest-mass density and increasing mass, simply because their size is also growing. This is a purely relativistic effect, absent in the Newtonian model of self-gravitating Keplerian disks \cite{mmp}. Sample graphs illustrating the dependence of the torus proper volume on the total ADM mass of the system and the outer equatorial circumferential radius of the disk are shown in Figs.\ \ref{madm-vol_m015_log_log_g43} and \ref{rc2-vol_m015_log_log_g43}, respectively. The proper volume of the torus is computed as
\begin{equation}
\label{volume}
V = 2 \pi \int dr \int d \theta r^2 \sin \theta \psi^6 e^{2 q},
\end{equation}
where the integrals are taken over the torus region. The graphs plotted in Figs.\ \ref{madm-vol_m015_log_log_g43} and \ref{rc2-vol_m015_log_log_g43} correspond to the data from Figs.\ \ref{biff_r2_100_g53} and \ref{biff_r2_10^15_g53} for the polytropic exponent $\Gamma = 4/3$. Solutions with the largest $\rho_\mathrm{max}$ were marked with triangles. The rapid growth of the torus volume coincides approximately with the decrease of the maximum density $\rho_\mathrm{max}$. Note that the proper volume of the torus can grow by more than 2 orders of magnitude in a series of models with fixed coordinate inner and outer radii. Further details of Figs.\ \ref{madm-vol_m015_log_log_g43} and \ref{rc2-vol_m015_log_log_g43} will be discussed in Sec.\ \ref{pappus_guldinus}.

It should be emphasised that the observed bifurcation cannot be fully explained by the lack of the geometric character of the radii $r_1$ and $r_2$. In Figure \ref{biff_rc075_m0_005} we plot the maximal density $\rho_\mathrm{max}$ versus the asymptotic mass $M_\mathrm{ADM}$ for a collection of solutions with $m = 1$, $a = 0$, $\Gamma = 4/3$ and fixed circumferential inner and outer radii of the torus: $r_\mathrm{c,1} \approx 100$ and $r_\mathrm{c,2} \approx 200$. Note that the maximal density $\rho_\mathrm{max}$ is not monotonic with respect to $M_\mathrm{ADM}$, and the bifurcation is still present. The dependence of the torus proper volume $V$ on the mass $M_\mathrm{ADM}$ for the data from Fig.\ \ref{biff_rc075_m0_005} is shown in Fig.\ \ref{vol_rc075_m0_005_log_log}. Not surprisingly, $V$ increases with $M_\mathrm{ADM}$, despite fixed $r_\mathrm{c,1}$ and $r_\mathrm{c,2}$. It can be checked that for these data $M_\mathrm{BH}$ also grows with $M_\mathrm{ADM}$, from the value $M_\mathrm{BH} \approx 1.003$ to $3.153$.

\section{Breaking of the Pappus-Guldinus rule}
\label{pappus_guldinus}

\epigraph{
``And I also met a certain number of professors there. [\dots] and the other explained to me that inside the globe there was another globe much bigger than the outer one.''}{\textit{Jaroslav Ha\v{s}ek}\\ \textit{The good soldier \v{S}vejk and his fortunes in the world war, Chapter 4.\ \v{S}vejk thrown out of the lunatic asylum}}

\begin{figure}
\includegraphics[width=\columnwidth]{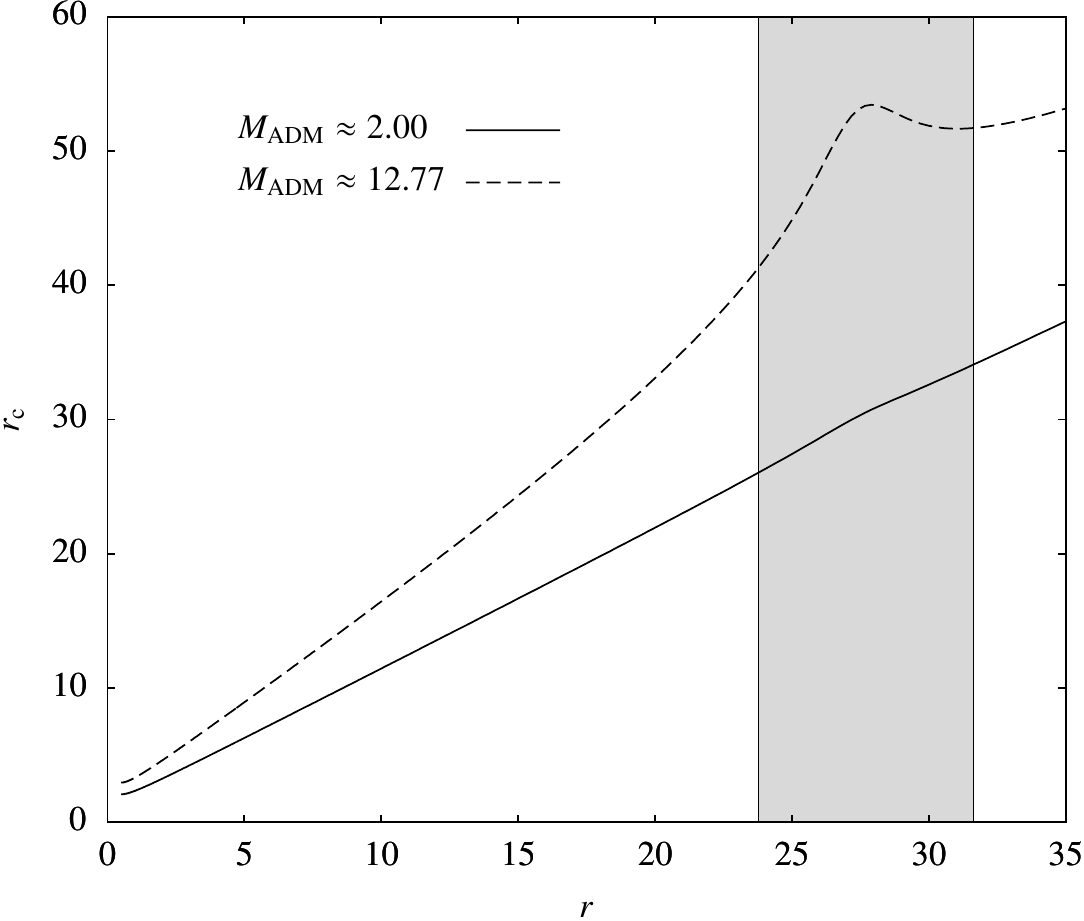}
\caption{\label{rc_g43} The circumferential radius $r_\textrm{c}$ versus the coordinate radius $r$ for $M_\textrm{ADM} \approx 2.00$ (solid line) and $M_\textrm{ADM} \approx 12.77$ (dashed line). Other parameters of the solutions are: $r_1 \approx 23.7$, $r_2 \approx 31.6$, $a=0$, $m=1$, and $\Gamma=4/3$. The area between $r_1$ and $r_2$ is marked in grey.}
\end{figure}

\begin{figure}
\includegraphics[width=\columnwidth]{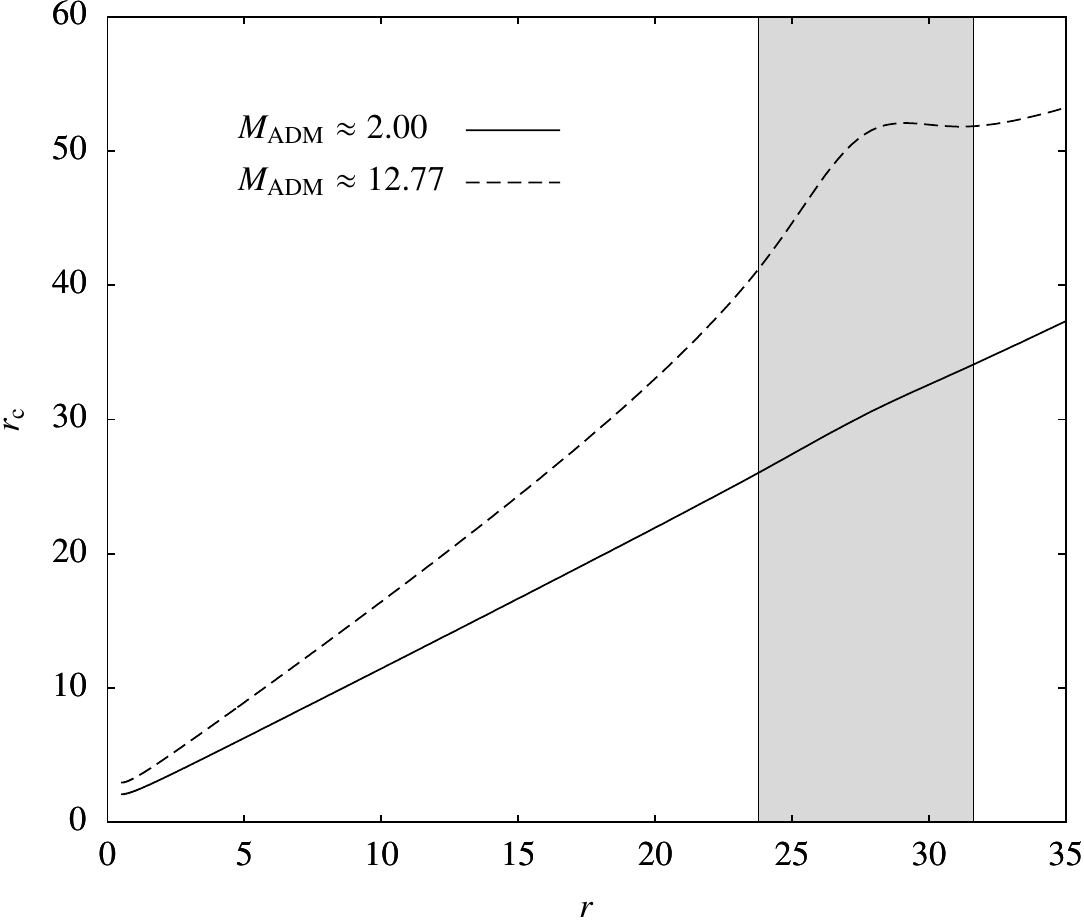}
\caption{\label{rc_g53} The same as in Fig.\ \ref{rc_g43}. Geometric parameters of the solutions are: $r_1 \approx 23.7$, $r_2 \approx 31.6$, $a=0$, $m=1$, and $\Gamma=5/3$. Again, the area between $r_1$ and $r_2$ is marked in grey.}
\end{figure}

\begin{figure}
\includegraphics[width=\columnwidth]{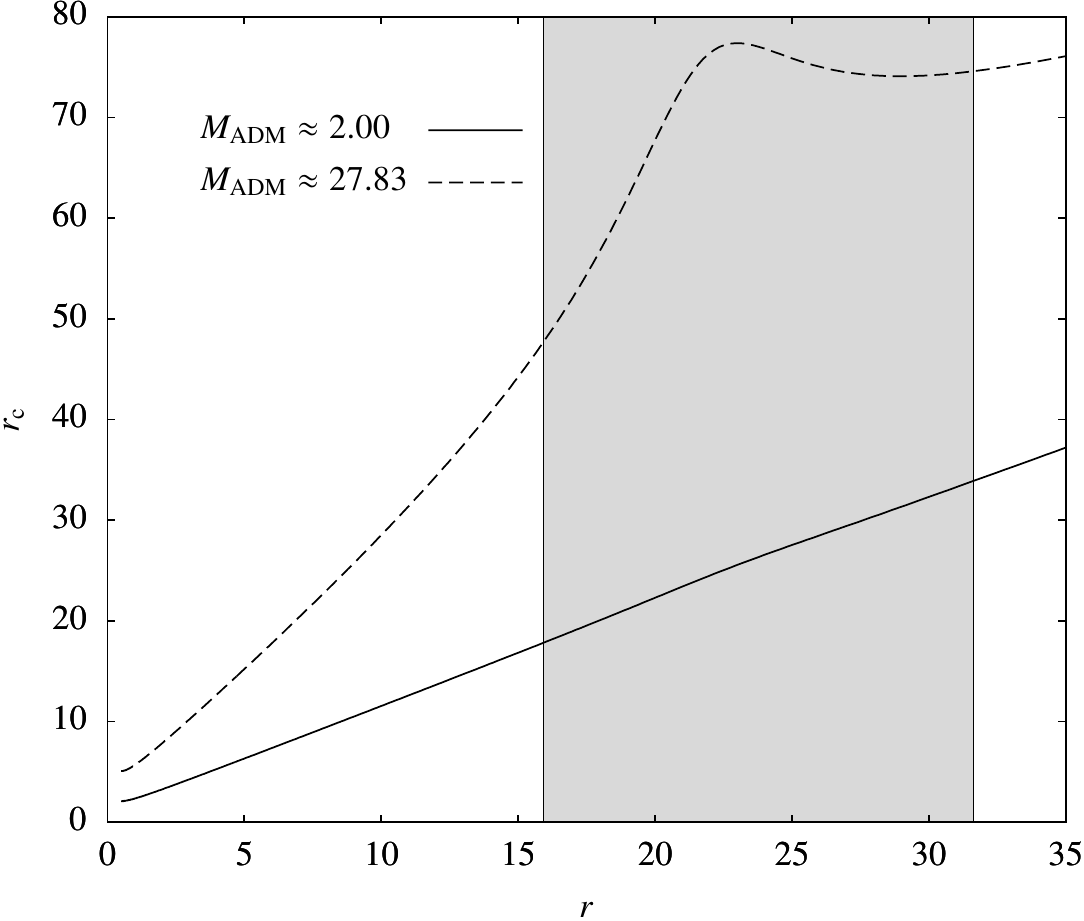}
\caption{\label{rc_r05_g43} The same as in Fig.\ \ref{rc_g43}. Geometric parameters of the solutions are: $r_1 \approx 15.8$, $r_2 \approx 31.6$, $a=0$, $m=1$, and $\Gamma=4/3$. Again, the area between $r_1$ and $r_2$ is marked in grey.}
\end{figure}

\begin{figure}
\includegraphics[width=\columnwidth]{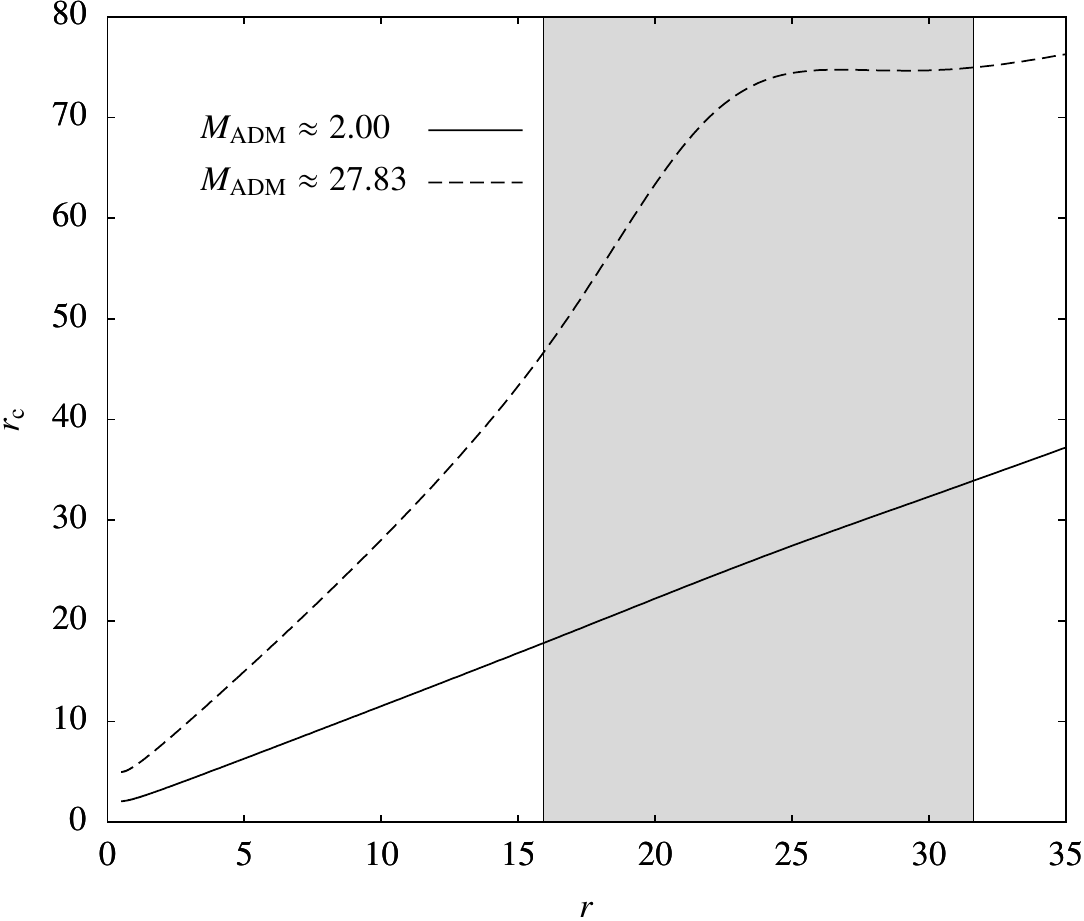}
\caption{\label{rc_r05_g53} The same as in Fig.\ \ref{rc_g43}. Geometric parameters of the solutions are: $r_1 \approx 15.8$, $r_2 \approx 31.6$, $a=0$, $m=1$, and $\Gamma=5/3$. Again, the area between $r_1$ and $r_2$ is marked in grey.}
\end{figure}

\begin{figure}
\includegraphics[width=\columnwidth]{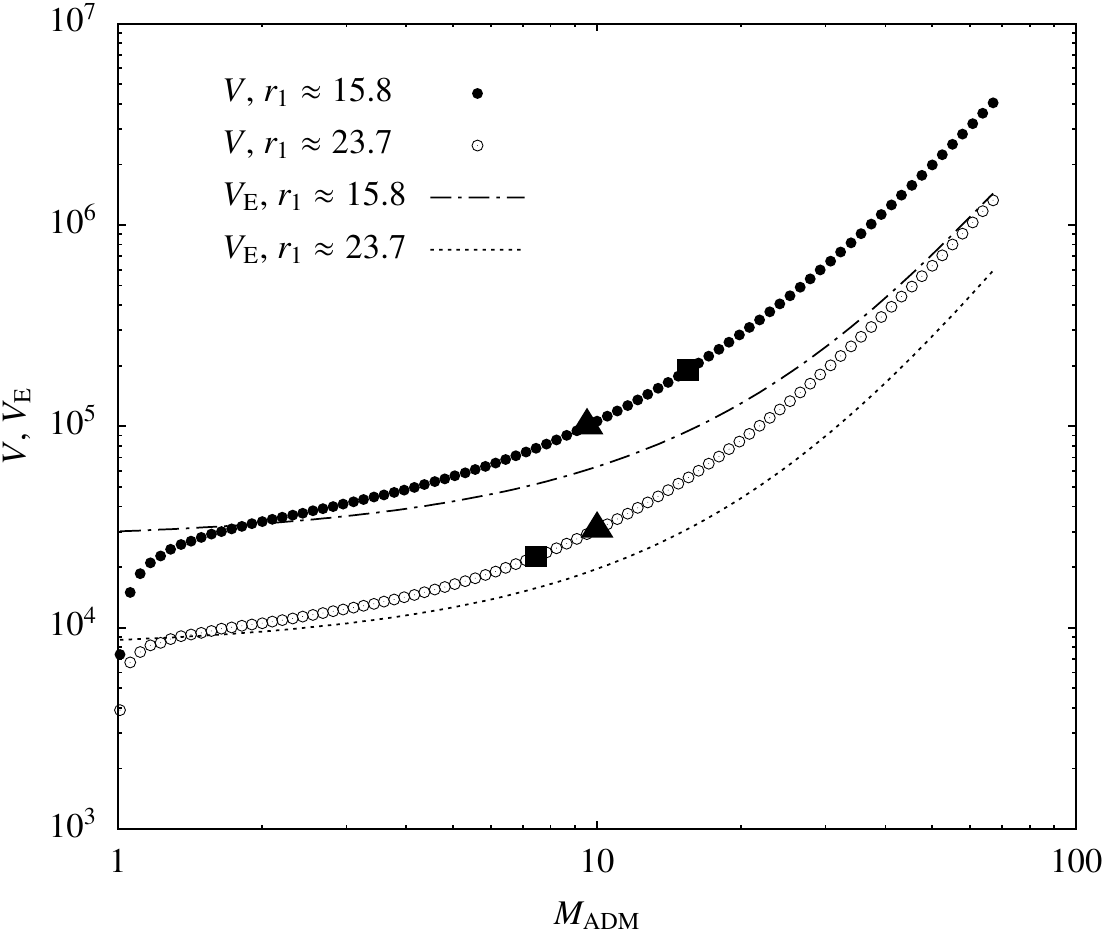}
\caption{\label{madm-vol_m015_log_log_g43} The proper volume of the torus $V$ and the Euclidean volume $V_\mathrm{E}$ versus the asymptotic mass $M_\textrm{ADM}$ for two classes of solutions with $r_1 \approx 15.8$ and $r_1 \approx 23.7$. Other parameters of the solutions are: $r_2 \approx 31.6$, $a=0$, $m=1$, and $\Gamma=4/3$. The Euclidean volume $V_\mathrm{E}$ is computed according to Eq.\ (\ref{volume_euclid}) (see the discussion in text). Squares represent solutions with the lowest mass for which a nonmonotonic behavior of $r_\textrm{c}$ with respect to $r$ has been observed. Triangles denote solutions with maximal values of  $\rho_\textrm{max}$.}
\end{figure}

\begin{figure}
\includegraphics[width=\columnwidth]{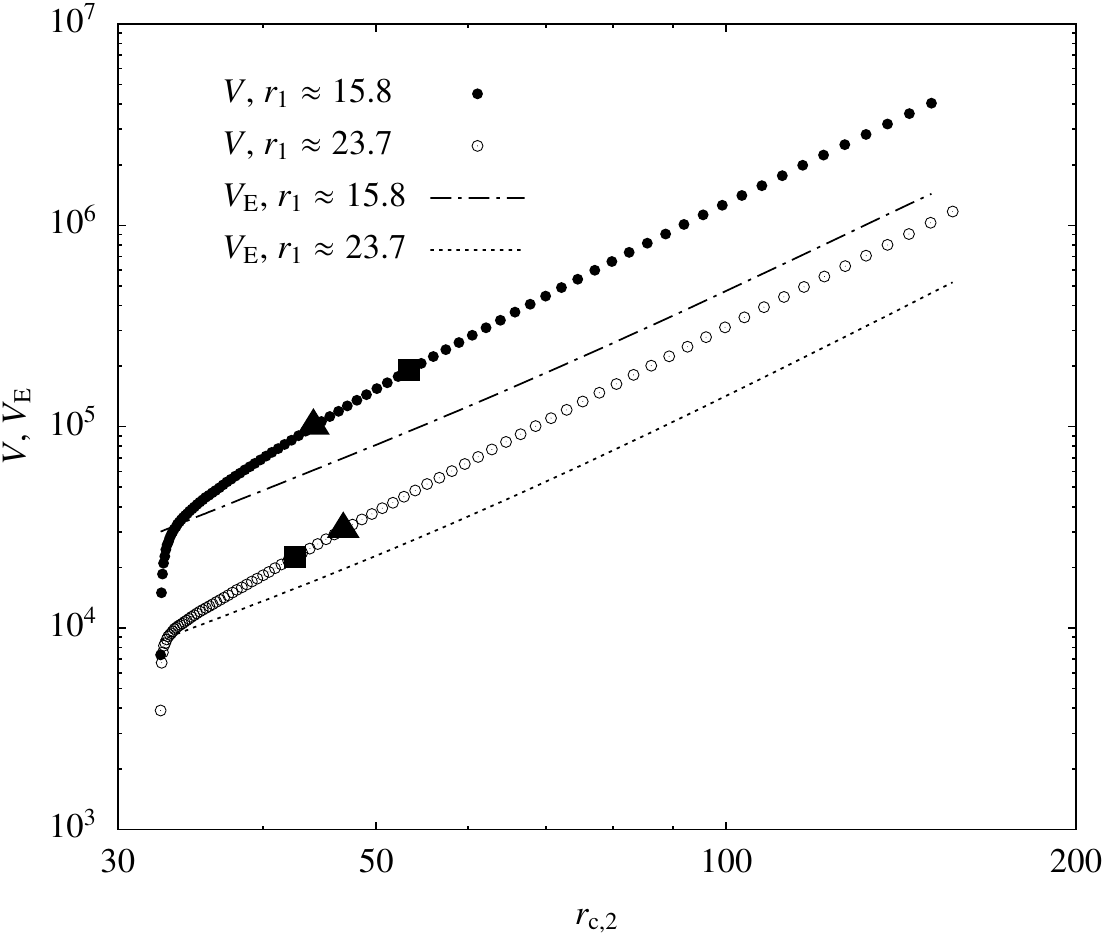}
\caption{\label{rc2-vol_m015_log_log_g43} The same data as in Fig.\ \ref{madm-vol_m015_log_log_g43}. The volumes $V$ and $V_\mathrm{E}$ are plotted against the outer equatorial circumferential radius of the torus $r_{\textrm{c},2}$. Both volumes grow asymptotically as $r_\mathrm{c,1}^3$.}
\end{figure}

\begin{figure}
\includegraphics[width=\columnwidth]{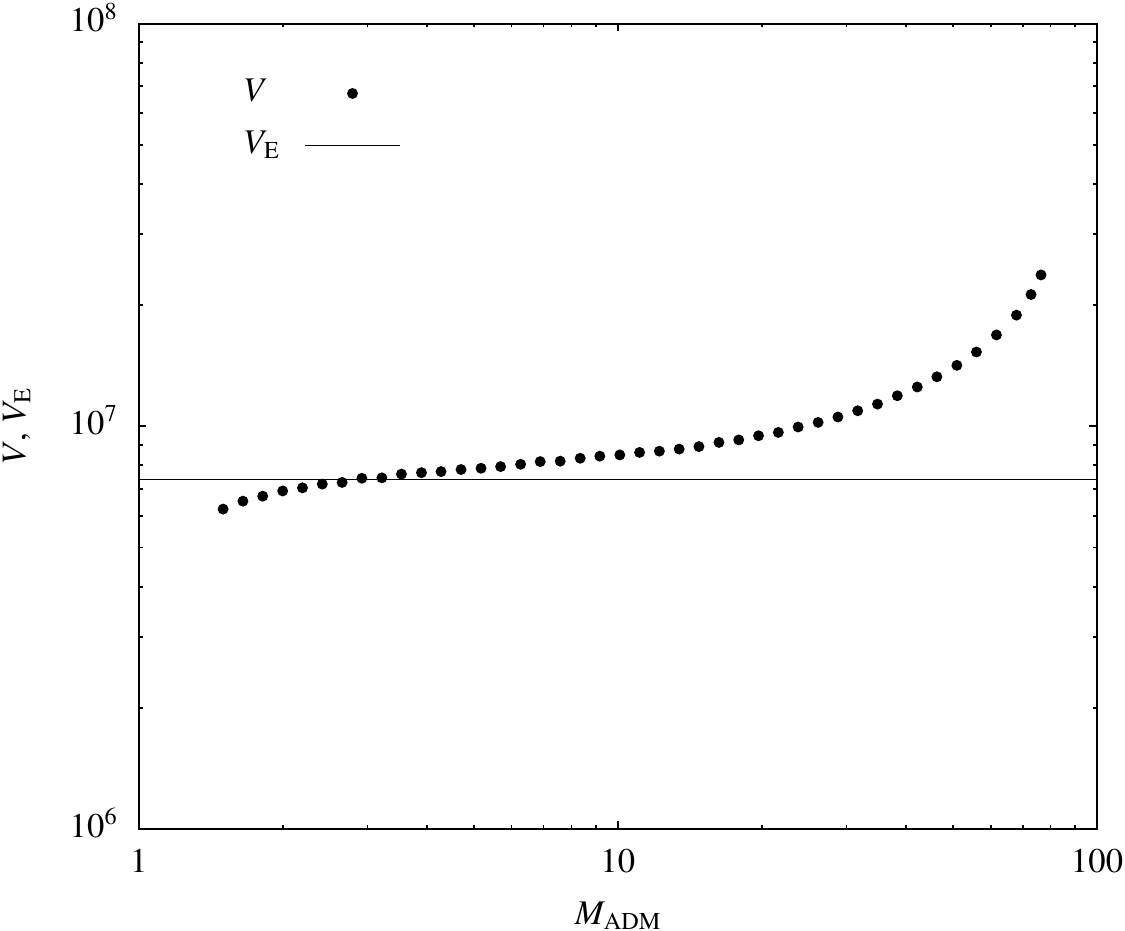}
\caption{\label{vol_rc075_m0_005_log_log} The proper volume of the torus $V$ and the Euclidean volume $V_\mathrm{E}$ versus the asymptotic mass $M_\textrm{ADM}$ for solutions with $r_{\textrm{c},1} \approx 100$, $r_{\textrm{c},2} \approx 200$, $a=0$, $m=1$, $\Gamma=4/3$. The Euclidean volume $V_\mathrm{E}$ is computed according to Eq.\ (\ref{volume_euclid}) (see the discussion in text).}
\end{figure}

It is common in mathematical relativity to define the geometric size of an axially symmetric body in terms of the largest circumference of a circle that can be embedded within this body. Equivalently, one can search for the maximum of $\eta_\mu \eta^\mu$ in the region occupied by the body \cite{khuri}. The result can be quite surprising, as the circle with the largest circumference does not have to be the outermost one. This is for instance the case of initial data investigated in \cite{kmmmx}. However, spotting such a behavior in more realistic examples is always surprising. In the context of self-gravitating stationary fluid disks, such a situation was observed by Labranche, Petroff, and Ansorg in \cite{labranche}. Figure 8 in \cite{labranche} shows the graph of the circumferential radius versus the proper (geodesic) radial distance at the equatorial plane of self-gravitating, rigidly rotating perfect fluid tori with a strange-matter equation of state \cite{gourgoulhon} and no black hole in the center. The dependence of $r_\mathrm{c}$ on the radial proper distance is not monotonic, and there is a clear local maximum of the circumferential radius. This figure is reproduced in \cite[][p.\ 148]{figuresequilibrium}, with added vertical lines marking the locations of the inner and outer edges of the tori. Clearly, the local maxima of the circumferential radius occur within the tori, meaning that the circles with the largest circumferences are not the outermost ones.

In Figures \ref{rc_g43}--\ref{rc_r05_g53} we show a similar behavior for our solutions. Since in our case a black hole is present, we simply plot the relation between the circumferential radius $r_\mathrm{c}$ and the coordinate radius $r$ at the equatorial plane. For clarity, the region between $r_1$ and $r_2$ is marked in grey.

The above behavior suggests that the volume of massive disks could be much larger than one could expect basing on the values of the outer circumferential radius of the torus. Because of the axially symmetric context, we will refer to this effect as breaking of the Pappus-Guldinus rule (theorem). In Euclidean geometry there are two famous theorems on solids of revolution attributed to Pappus of Alexandria and Paul Guldin \cite{pappus,guldin}. The first yields the area of the surface of a solid of revolution. The second, which is of our interest here, states that the volume of a solid of revolution, obtained by rotating a plane figure $F$ with a surface $A$ about an external axis, can be expressed as
\begin{equation}
\label{pappus}
V_\mathrm{E} = A d,
\end{equation}
where $d$ is the distance traveled by the geometric center of $F$. For instance, in Euclidean geometry the volume of a proper torus (obtained by rotating a circle about an external axis) can be expressed as
\begin{equation}
\label{volume_euclid}
V_\mathrm{E} = 2 \pi \frac{r_1 + r_2}{2} \pi \left( \frac{r_2 - r_1}{2} \right)^2 = \frac{\pi^2}{4} (r_1 + r_2) (r_2 - r_1)^2,
\end{equation}
where $r_1$ and $r_2$ denote the torus inner and outer radius, respectively.

In the examples shown in Figs.\ \ref{madm-vol_m015_log_log_g43} and \ref{rc2-vol_m015_log_log_g43} we compute the Euclidean volume $V_\mathrm{E}$ given by Eq.\  (\ref{volume_euclid}), assuming in place of $r_1$ and $r_2$ the inner and outer circumferential radii $r_\mathrm{c,1}$ and $r_\mathrm{c,2}$ of the obtained toroidal fluid configuration, and then compare the result with the true proper volume $V$ given by Eq.\ (\ref{volume}). Figures \ref{madm-vol_m015_log_log_g43} and \ref{rc2-vol_m015_log_log_g43} show both volumes $V$ and $V_\mathrm{E}$ plotted against the mass $M_\mathrm{ADM}$ and the outer circumferential radius $r_\mathrm{c,2}$, respectively. The logic behind this comparison is as follows. Firstly, we observe that the meridional cross-sections of massive tori are approximately circular. Secondly, we imagine that for some reason an observer gathers information about the size of the torus in terms of its ``external'' geometric characteristics---for instance the inner and outer circumferential radii. In this case expression (\ref{volume_euclid}) can serve as a natural scale of the volume associated with the system. The results shown in Figs.\ \ref{madm-vol_m015_log_log_g43} and \ref{rc2-vol_m015_log_log_g43} mean that the massive tori are in fact much larger (in the sense of the proper volume) than they appear from the outside. The proper volume $V$ is smaller than $V_\mathrm{E}$ given by Eq.\ (\ref{volume_euclid}) only for light tori, which are also geometrically thin.

Perhaps the most clear illustration of the above effect is given in Fig.\ \ref{vol_rc075_m0_005_log_log}, where we plot the proper volume $V$ for a collection of tori with fixed circumferential radii $r_\mathrm{c,1}$ and $r_\mathrm{c,2}$ and a growing mass $M_\mathrm{ADM}$. A fixed value of $V_\mathrm{E}$ computed according to Eq.\ \ref{volume_euclid} is denoted with a horizontal line.

Strictly speaking, the term ``breaking of the Pappus-Guldinus rule'' should apply to a deviation from Eq.\ (\ref{pappus}) rather than Eq.\ (\ref{volume_euclid}). In other words, one could compute the exact proper area of the meridional cross-section of a toroidal body, define the geometric center of this cross-section (this is of course ambiguous), and compare the proper volume $V$ of the whole body with $V_\mathrm{E}$ given by Eq.\ (\ref{volume_euclid}). On the other hand, for our tori, the proper area of the meridional cross-section is not a measurable quantity, whereas circumferential radii $r_\mathrm{c,1}$ and $r_\mathrm{c,2}$ could, in principle, be measured. A comparison of $V$ with $V_\mathrm{E}$ given by Eq.\ (\ref{volume_euclid}) serves as a poor-man approach to the breaking of the Pappus-Guldinus theorem. 

The fact that the volume of bodies in General Relativity can be much larger than it is suggested by their external size measures is a well-known result in spherical symmetry \cite{bmm, kmmmx}. As far as we know, in spherical symmetry this effect is present for bodies contained within the apparent horizon. In contrast to that, the tori discussed in this paper are located outside the horizon.

In Figures \ref{madm-vol_m015_log_log_g43} and \ref{rc2-vol_m015_log_log_g43}, in addition to triangles marking the solutions characterized by the largest value of $\rho_\mathrm{max}$, we have also drawn small squares denoting the least massive solutions for which the relation between the circumferential radius $r_\mathrm{c}$ and the coordinate radius $r$ at the equatorial plane ceases to be monotonic. There seems to be no direct correlation between maximizing $\rho_\mathrm{max}$ and nonmonotonicity of $r_\mathrm{c}$ with respect to $r$.

\section{Ergoregions}
\label{secergo}

An ergoregion (or traditionally an ergosphere) is defined as a region outside the black hole horizon, where the Killing vector $\xi^\mu$, which is asymptotically timelike (i.e., $\xi_\mu \xi^\mu < 0$), becomes spacelike ($\xi_\mu \xi^\mu > 0$). For the Killing vector $\xi^\mu = (1,0,0,0)$ and metrics (\ref{generalmetric}) or (\ref{isotropic}) this actually means that
\begin{equation}
g_{\mu \nu} \xi^{\mu} \xi^{\nu} = g_{t t} = -\alpha^2 +\psi^4 r^2 \sin ^2 \theta \beta^2 > 0.
\end{equation}
A surface defined by the condition $\xi_\mu \xi^\mu = 0$ is usually called an ergosurface.

If the disk (torus) is relatively light, the only ergoregion in the black hole--torus system is connected with the central black hole, and it deviates very slightly from the ergoregion of the Kerr black hole. This can change for massive tori. In \cite{ansorg:2006} Ansorg and Petroff gave numerical examples of constant energy density tori, rigidly rotating around spinning black holes, with ergoregions of different topologies. A massive torus can create its own toroidal ergoregion, in addition to the one connected with the central black hole. This toroidal ergoregion grows with the increasing mass of the torus, and it can encompass the entire torus. If the torus is sufficiently massive and compact, the two ergoregions---that of the torus and of the black hole---can merge. We recover this behavior in our models of differentially rotating, polytropic tori. To the best of our knowledge, very few authors have investigated this or similar effects so far. In \cite{chrusciel} Chru\'{s}ciel, Greuel, Meinel, and Szybka analyzed the regularity of the ergosurface in the vacuum region (the analysis is based on the Ernst equation). In particular, motivated by Ansorg and Petroff results, they discuss the coalescence of two ergosurfaces. A toroidal ergoregion around a black hole was also observed for configurations of selfgravitating scalar fields in \cite{Herdeiro15}.
 
Parameters of our solutions, illustrating different possible configurations of ergoregions, are collected in Table \ref{tabErgReg}. All these solutions have been obtained assuming the polytropic equation of state with the exponent $\Gamma = 4/3$ and the  Keplerian rotation law (\ref{keplerian_rl}). We set the black hole parameters $m = 1$ and $a = 0.96$.

\begin{figure}
\includegraphics[width=\columnwidth]{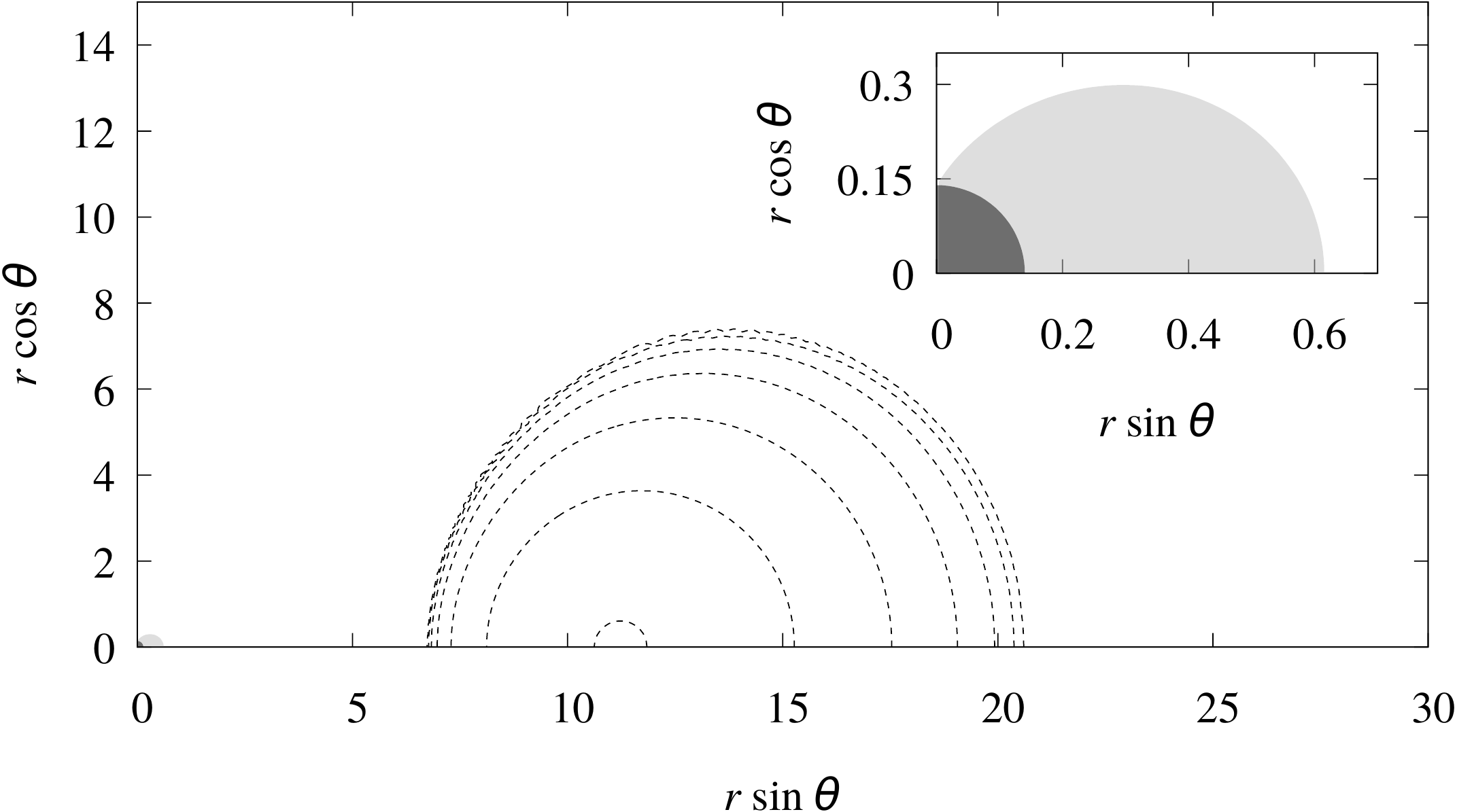}
\caption{\label{ErgRegBHonly} Solution E1 from Table \ref{tabErgReg}. The ergoregion is marked in grey. Black color marks the region inside the horizon. Broken density isolines correspond to $\rho = 8 \times 10^i$, $i = -10, -9, \dots, -4$.}
\end{figure}

\begin{figure}
\includegraphics[width=\columnwidth]{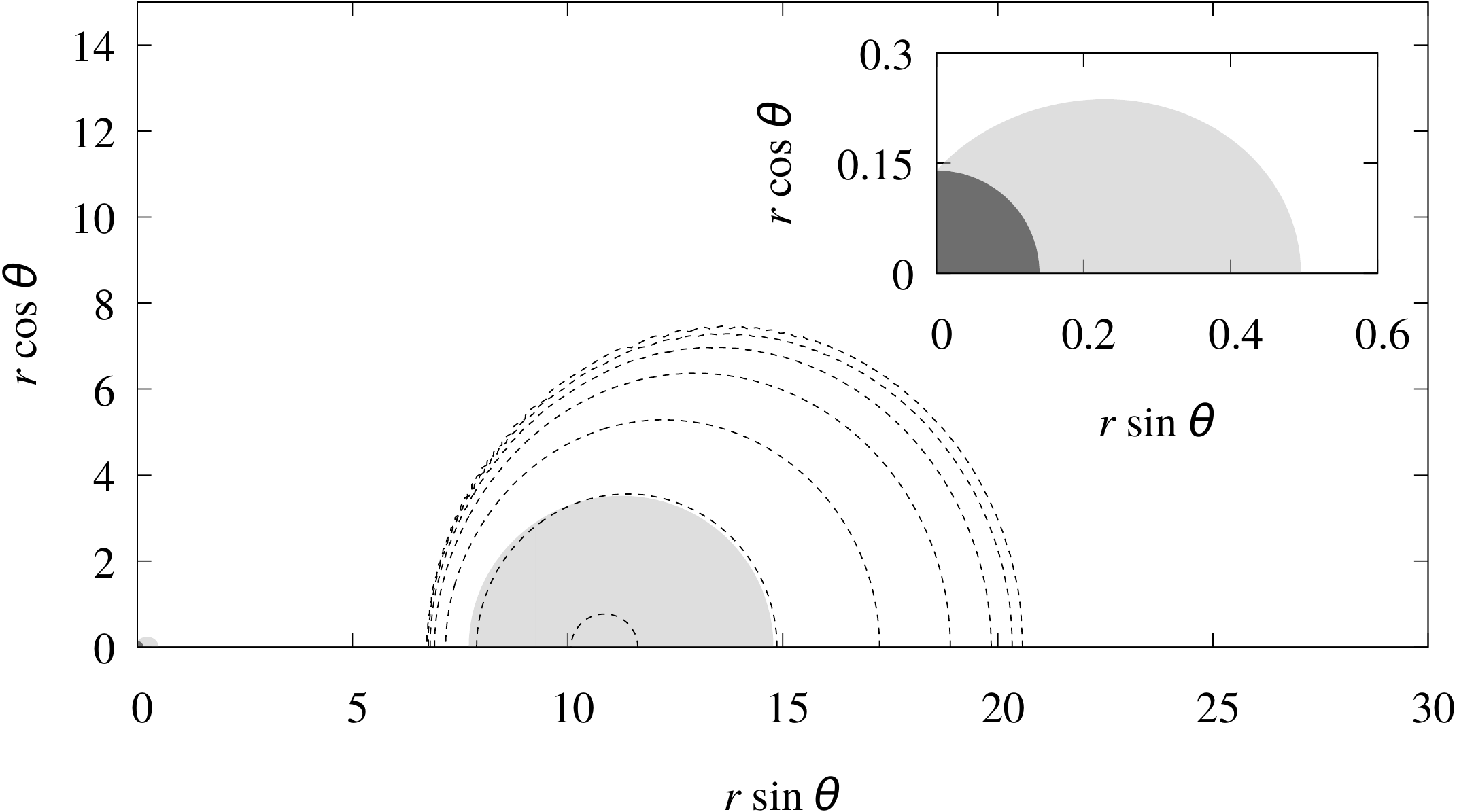}
\caption{\label{ErgRegInDisk} Solution E2 from Table \ref{tabErgReg}. The ergoregions are marked in grey. Black color marks the region inside the horizon. Broken density isolines correspond to $\rho = 6 \times 10^i$, $i = -10, -9, \dots, -4$.}
\end{figure}

\begin{figure}
\includegraphics[width=\columnwidth]{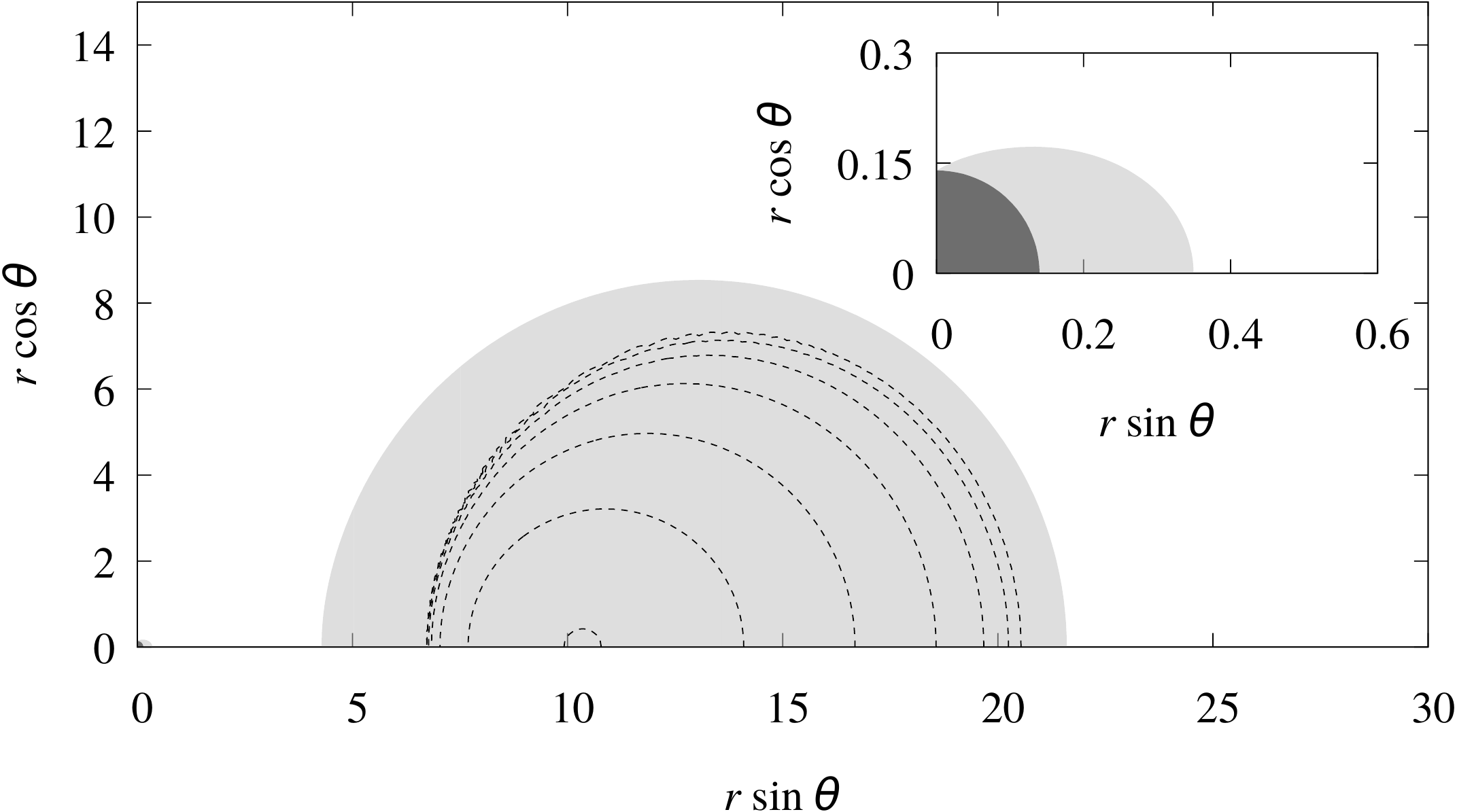}
\caption{\label{ErgRegOutsideDisk} Solution E3 from Table \ref{tabErgReg}. The ergoregions are marked in grey. Black color marks the region inside the horizon. Broken density isolines correspond to $\rho = 4 \times 10^i$, $i = -10, -9, \dots, -4$.}
\end{figure}

\begin{table*}
\caption{\label{tabErgReg} Solutions with different ergoregion configurations. Subseqent columns contain (from the left to the right): solution number, the total asymptotic mass $M_\mathrm{ADM}$, the black hole mass $M_\mathrm{BH}$, the inner radius of the torus $r_1$, the outer radius of the torus $r_2$,  the radius of the ergoregion corresponding to the black hole $r_\mathrm{ergBH}$, the inner radius of the ergoregion connected with the torus $r_\mathrm{ergT1}$, the outer radius of the ergoregion connected with the torus $r_\mathrm{ergT2}$, the maximal rest-mass density within the torus $\rho_\mathrm{max}$. All radii  $r_1$, $r_2$, $r_\mathrm{ergBH}$, $r_\mathrm{ergT1}$, and $r_\mathrm{ergT2}$ are defined as coordinate radii at the equatorial plane. 
The solutions were obtained assuming $m = 1$, $a = 0.96$, and $\Gamma = 4/3$.}
\begin{ruledtabular}
\begin{tabular}{c c c c c c c c c }
No. & $M_{\mathrm{ADM}}$ & $M_{\mathrm{BH}}$ & $r_1$ & $r_2$  & $r_{\mathrm{ergBH}}$ & $r_{\mathrm{ergT1}}$ & $r_{\mathrm{ergT2}}$ & $\rho_{\text{max}}$ \\
\hline
E1 & $8.00$ & $1.03$ & $6.7$ & $20.8$  & $0.62$ & $-$ & $-$  & $8.7 \times 10^{-4}$ \\
E2 & $12.00$ & $1.06$ & $6.7$ & $20.8$  & $0.49$ & $7.70$ & $14.8$ & $7.0 \times 10^{-4}$ \\
E3 & $20.00$ & $1.17$ & $6.7$ & $20.8$ & $0.35$ & $4.28$ & $21.6$  & $4.3 \times 10^{-4}$ \\
E4 & $4.00$ & $1.03$ & $2.0$ & $3.1$  & $-$ & $-$ & $4.4$  & $5.4 \times 10^{-2}$ \\
\end{tabular}
\end{ruledtabular}
\end{table*} 

\begin{table*}
\caption{\label{tabErgReg2} Maximal densities within the torus for solutions reported in Table \ref{tabErgReg}. The columns report rest-mass densities in $\mathrm{kg}/\mathrm{m}^3$, assuming the black hole mass parameter $m = (1, 10, 50, 10^6, 10^9) M_\odot$. Here $M_\odot =1.988\times 10^{30}\, \mathrm{kg} $.}
\begin{ruledtabular}
\begin{tabular}{c c c c c c}
No. & $\rho_\mathrm{max}(m = 1M_{\odot})$  & $\rho_\mathrm{max}(m = 10M_{\odot})$ & $\rho_\mathrm{max}(m = 50M_{\odot})$ & $\rho_\mathrm{max}(m = 10^6M_{\odot})$ & $\rho_\mathrm{max}(m = 10^9M_{\odot})$ \\
\hline
E1 & $5.4\times 10^{17}$ & $5.4 \times 10^{15}$ & $2.2 \times 10^{14}$ & $5.4 \times 10^{5}$ & $5.4 \times 10^{-1}$ \\
E2 & $4.4 \times 10^{17}$ & $4.4 \times 10^{15}$ & $1.7 \times 10^{14}$ & $4.4 \times 10^{5}$ & $4.4 \times 10^{-1}$ \\
E3 & $2.6 \times 10^{17}$ & $2.6 \times 10^{15}$ & $1.1 \times 10^{14}$ & $2.6 \times 10^{5}$ & $2.6 \times 10^{-1}$ \\
E4 & $3.4 \times 10^{19}$ & $3.4 \times 10^{17}$ & $1.3 \times 10^{16}$ & $3.4 \times 10^{7}$ & $3.4 \times 10^{1}$ \\
\end{tabular}
\end{ruledtabular}
\end{table*}

Solution E1 is depicted in Fig.\ \ref{ErgRegBHonly}. It corresponds to a standard case with a single Kerr-like ergoregion connected with the rotating black hole. Here the total asymptotic mass is already large, $M_\mathrm{ADM} = 8.00$. The mass of the black hole reads $M_\mathrm{BH} = 1.03$. Increasing the total ADM mass to $M_\mathrm{ADM} = 12.00$, we obtained a configuration (labeled as E2 in Table \ref{tabErgReg}) with two disconnected ergoregions---one surrounding the black hole, and a toroidal one, corresponding to the torus. Both these ergoregrions are shown in Fig.\ \ref{ErgRegInDisk}.

Increasing the ADM mass even further, to $M_\mathrm{ADM} = 20.00$, we obtained solution E3 with the toroidal component of the ergoregion encompassing the matter torus; this solution is shown in Fig.\ \ref{ErgRegOutsideDisk}.

\begin{figure}
\includegraphics[width=\columnwidth]{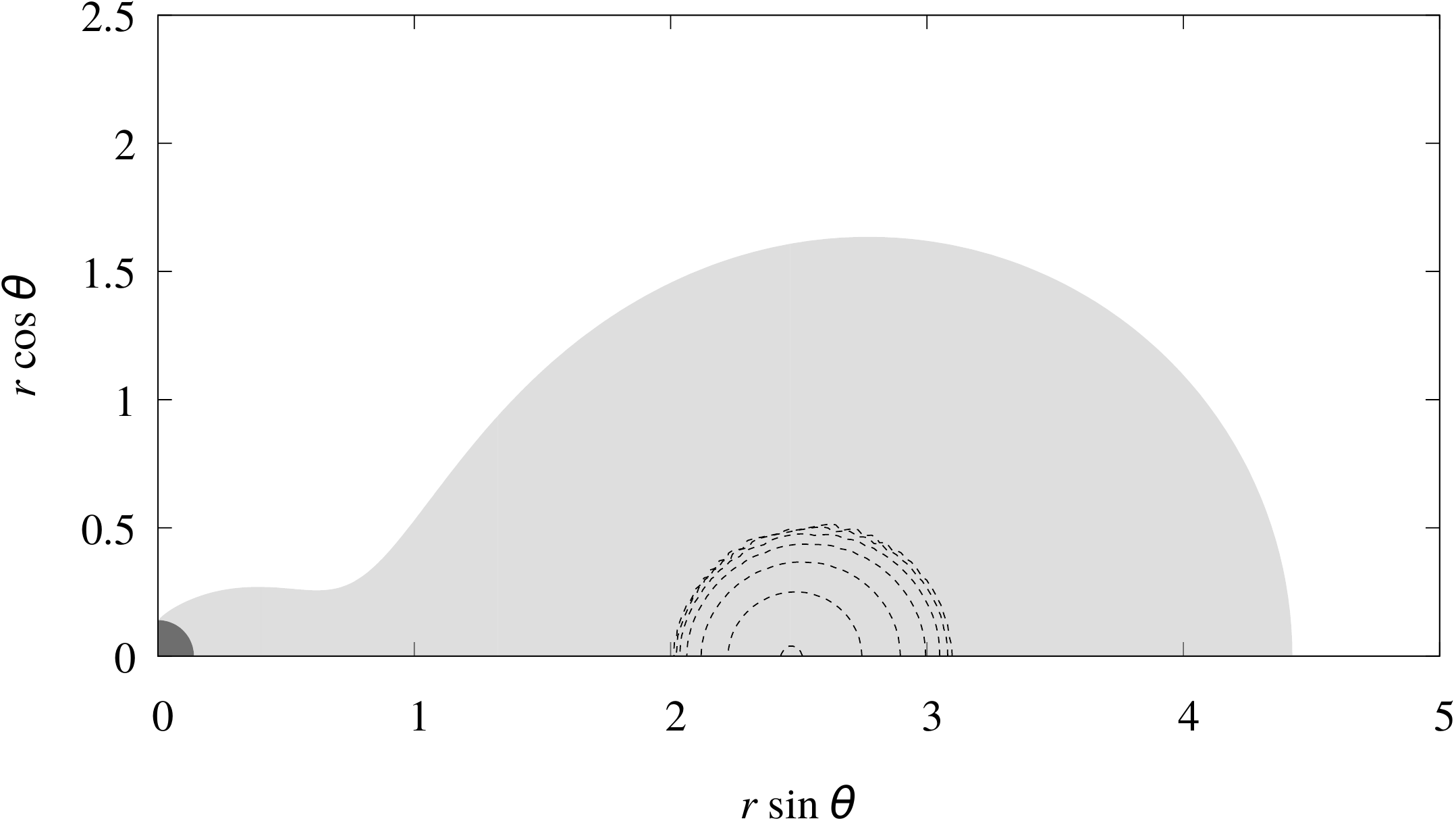}
\caption{\label{ErgRegMerg1}
Solution E4 from Table \ref{tabErgReg}. The ergoregion is marked in grey. Black color marks the region inside the horizon. Broken density isolines correspond to $\rho = 5 \times 10^i$, $i = -8, -7, \dots, -2$.}
\end{figure}

Similarly to the solution discussed by Ansorg and Petroff in \cite{ansorg:2006}, it is also possible to obtain a stationary configuration with a single ergoregion encompassing both the black hole and the torus and the ergosurface of spherical topology. We obtain such a solution assuming an extremely compact configuration---the inner and outer coordinate radii of the torus read $r_1 = 2.0$ and $r_2 = 3.1$, respectively. This solution, denoted as E4 in Table \ref{tabErgReg}, is shown in Fig.\ \ref{ErgRegMerg1}.

All tori E1--E4 are dense. Translating the maximal densities of the tori reported in Table \ref{tabErgReg} into physical units (say SI units) is simple, but it requires breaking of the scaling symmetry inherent to the equations describing the stationary black hole--torus system. This can be done, for instance, by specifying the mass of the central black hole or the black hole mass parameter $m$. Sample values of the maximal densities of the tori expressed in $\mathrm{kg}/\mathrm{m}^3$ for $m = (1, 10, 50, 10^6, 10^9) M_\odot$, are collected in Table \ref{tabErgReg2}. For stellar mass black holes we get the maximal densities comparable to those occurring in the centers of neutron stars. For supermassive black holes this density is comparable to the density of the air.

It is possible to obtain massive tori with small densities, provided that their volumes are sufficiently large. Examples of such solutions were discussed in Sec.\ \ref{secbif}. These solutions can also exhibit disconnected ergoregions, similar to the ones shown in Figs.\ \ref{ErgRegInDisk} and \ref{ErgRegOutsideDisk}.

\section{Circular geodesics}
\label{circgeodesics}

\begin{figure}
\includegraphics[width=\columnwidth]{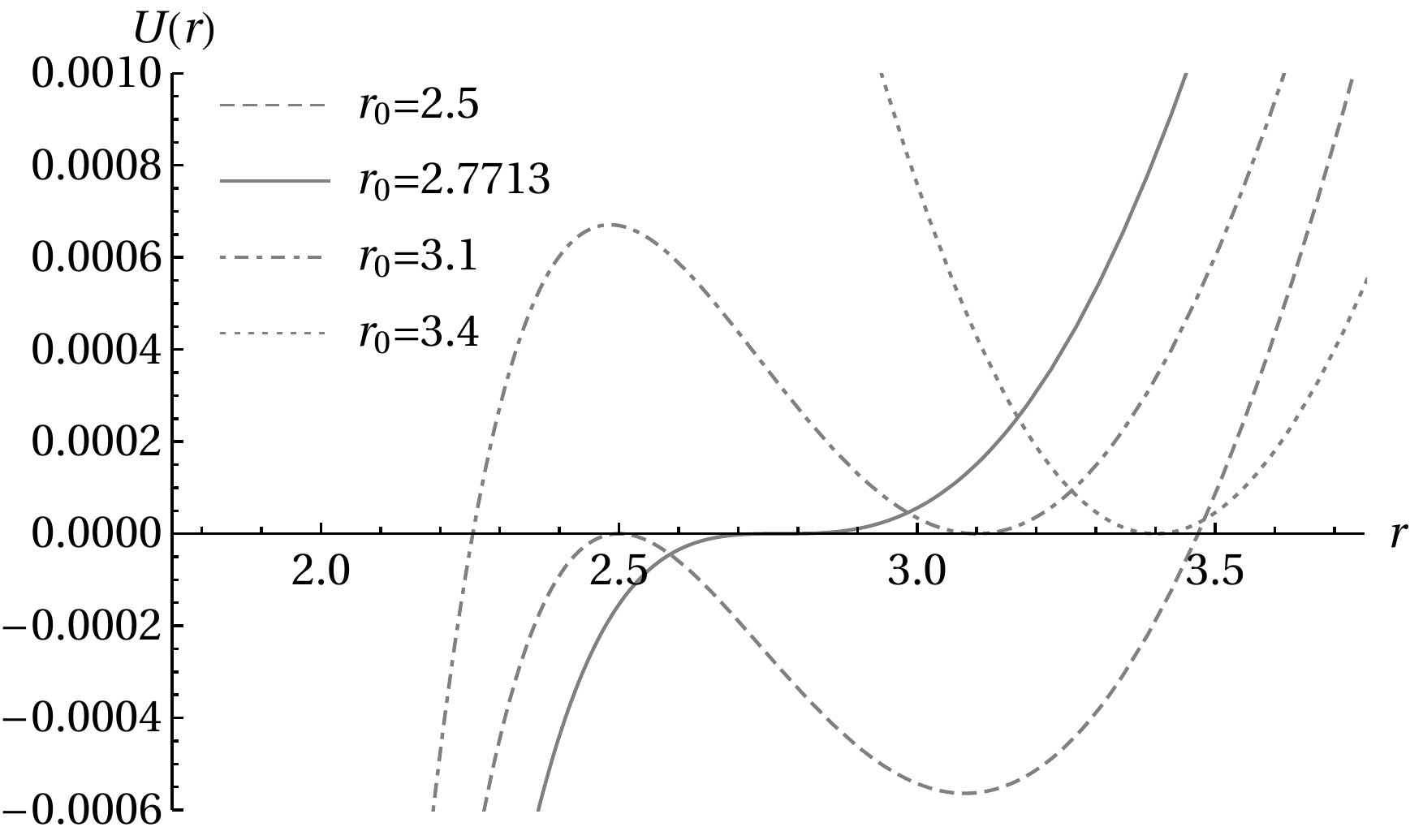}
\caption{\label{iscokerr}
The potential $U(r)$ for 4 circular prograde geodesics in the Kerr spacetime (in quasiisotropic coordinates) with $m = 1$ and $a = 0.6$. The ISCO is located at $r_0 \approx 2.7713$. The graphs correspond to 4 circular geodesics with radii $r_0 = 2.5, 2.7713, 3.1, 3.4$.}
\end{figure}

\begin{figure}
\includegraphics[width=\columnwidth]{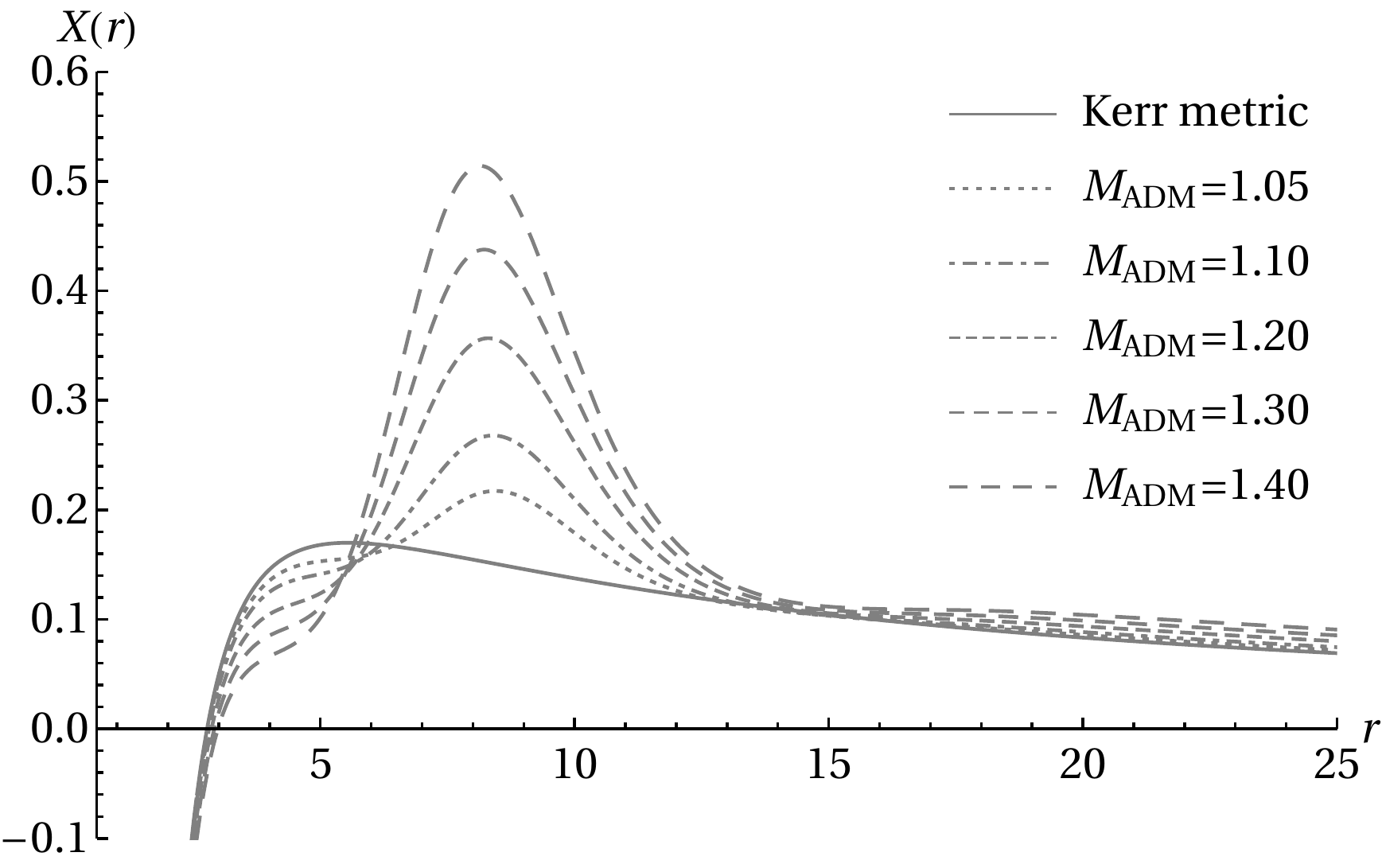}
\caption{\label{iscoa6XrBH}
The function $X(r)$ for the Kerr solution and five additional solutions (I1--I5) with $m = 1$, $a = 0.6$, $\Gamma=4/3$, $r_1 = 2.9$, $r_2 = 18.1$, and different asymptotic masses $M_\mathrm{ADM} = 1.05, 1.1, 1.2, 1.3, 1.4$.}
\end{figure}

\begin{figure}
\includegraphics[width=\columnwidth]{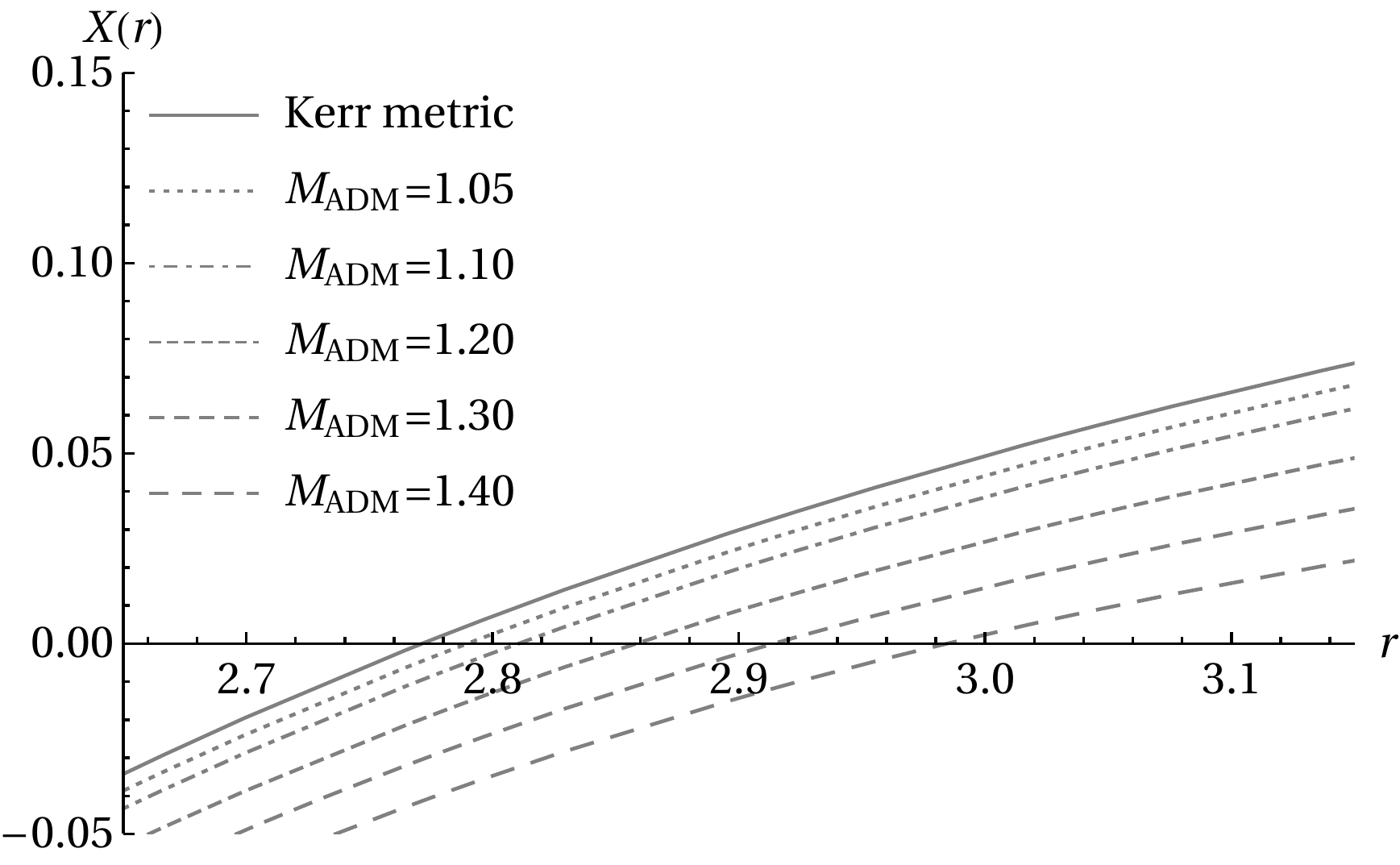}
\caption{\label{iscoa6Xr}
The same as in Fig.\ \ref{iscoa6XrBH}. The function $X(r)$ is plotted in the neighborhood of the ISCO.}
\end{figure}

\begin{figure}
\includegraphics[width=\columnwidth]{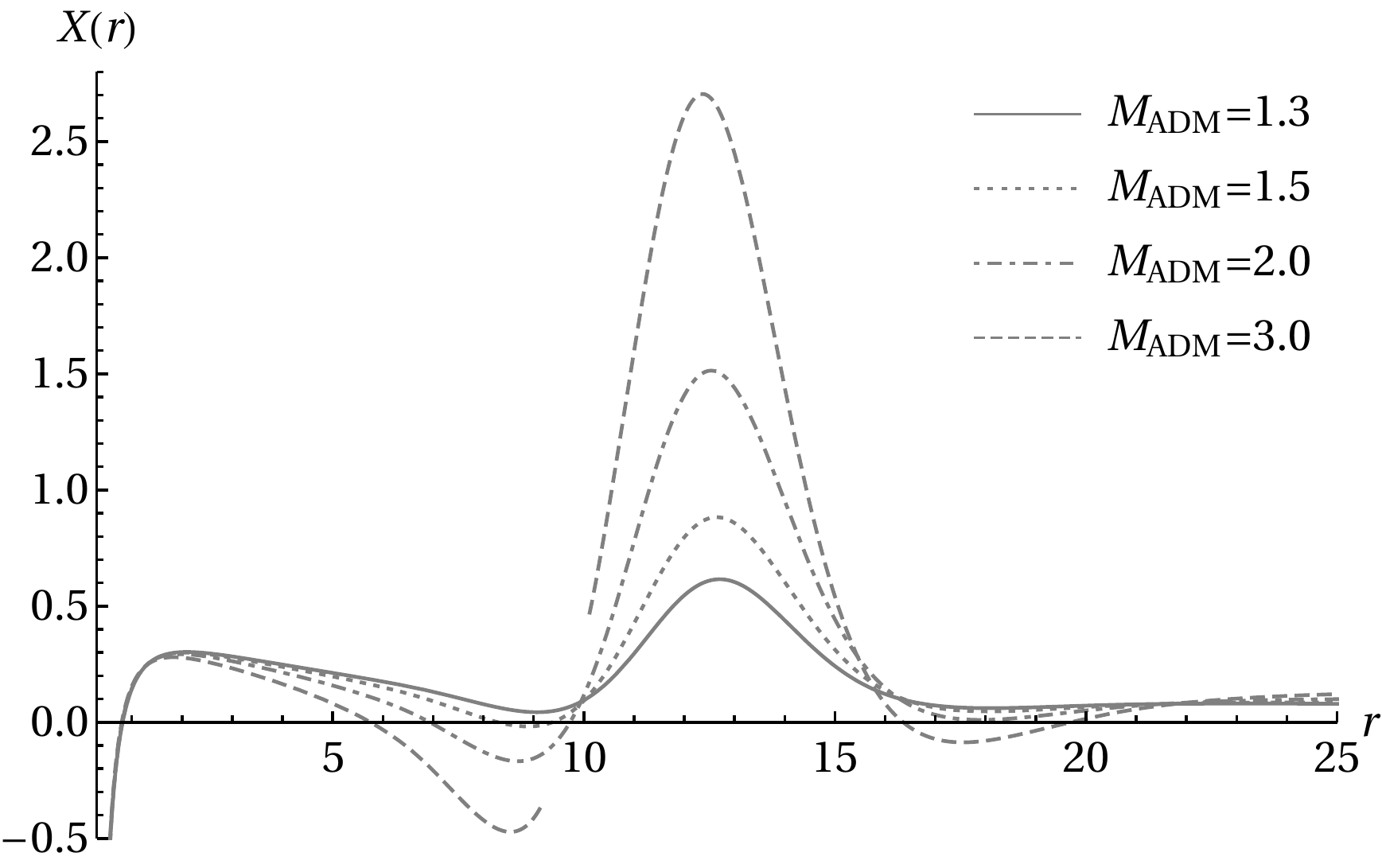}
\caption{\label{iscoa96Xr}
The function $X(r)$ for 4 solutions with $m = 1$, $a = 0.96$, $\Gamma=4/3$, $r_1 = 6.7$, $r_2 = 20.8$ and different asymptotic masses $M_\mathrm{ADM} = 1.3, 1.5, 2, 3$. There is an additional region with $X(r) < 0$ (unstable geodesic orbits) for $M_\mathrm{ADM} = 1.5$ and $2.0$. For $M_\mathrm{ADM} = 3$, there is a region within the torus, in which no circular geodesic orbits exist [it is depicted as a gap in the graph of $X(r)$].}
\end{figure}

\begin{figure}
\includegraphics[width=\columnwidth]{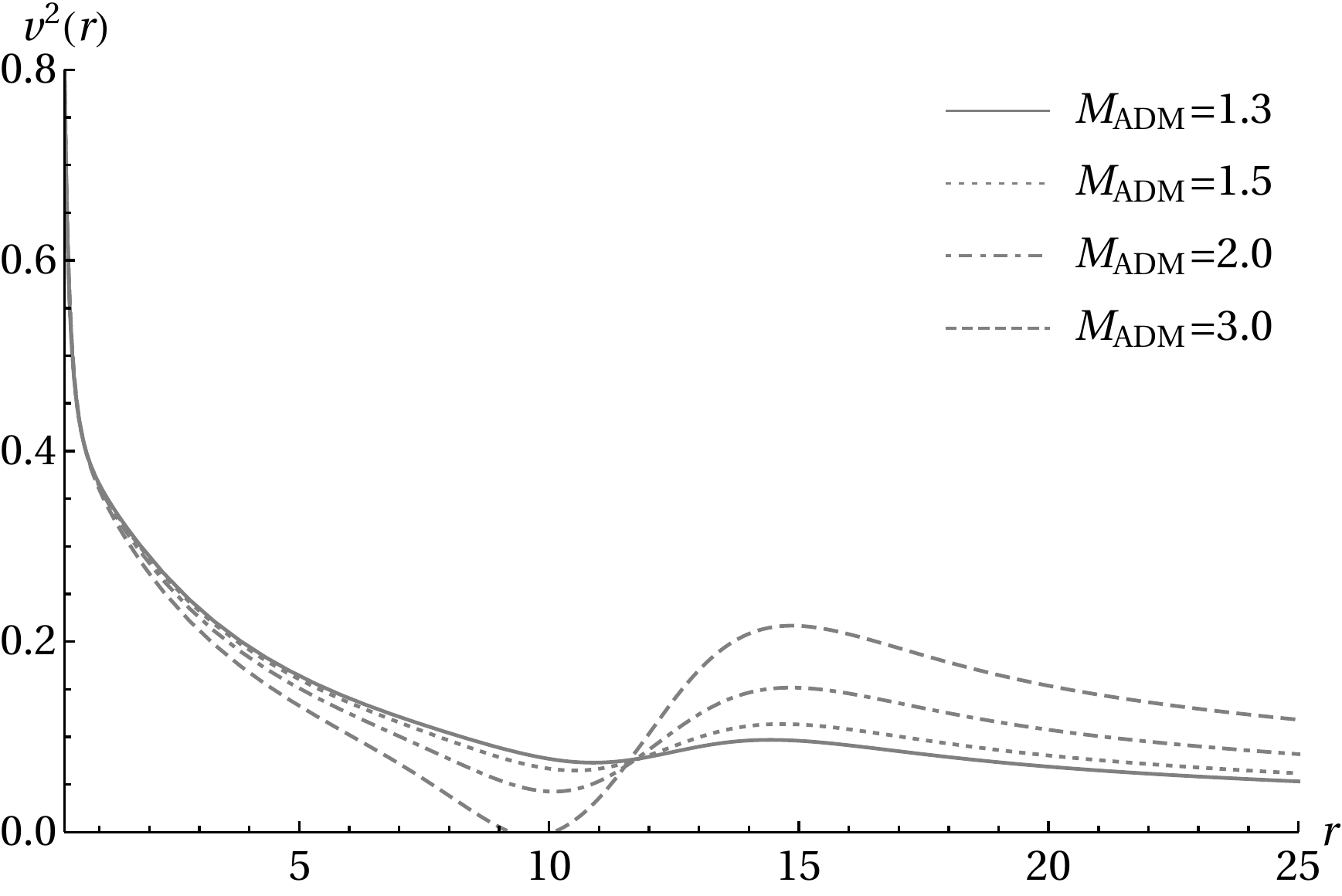}
\caption{\label{V2a96Xr}
The square of linear velocity $v^2$ associated with circular geodesics for 4 solutions with $m = 1$, $a = 0.96$, $\Gamma=4/3$, $r_1 = 6.7$, $r_2 = 20.8$, and different asymptotic masses $M_\mathrm{ADM} = 1.3, 1.5, 2, 3$.}
\end{figure}

\begin{figure}
\includegraphics[width=\columnwidth]{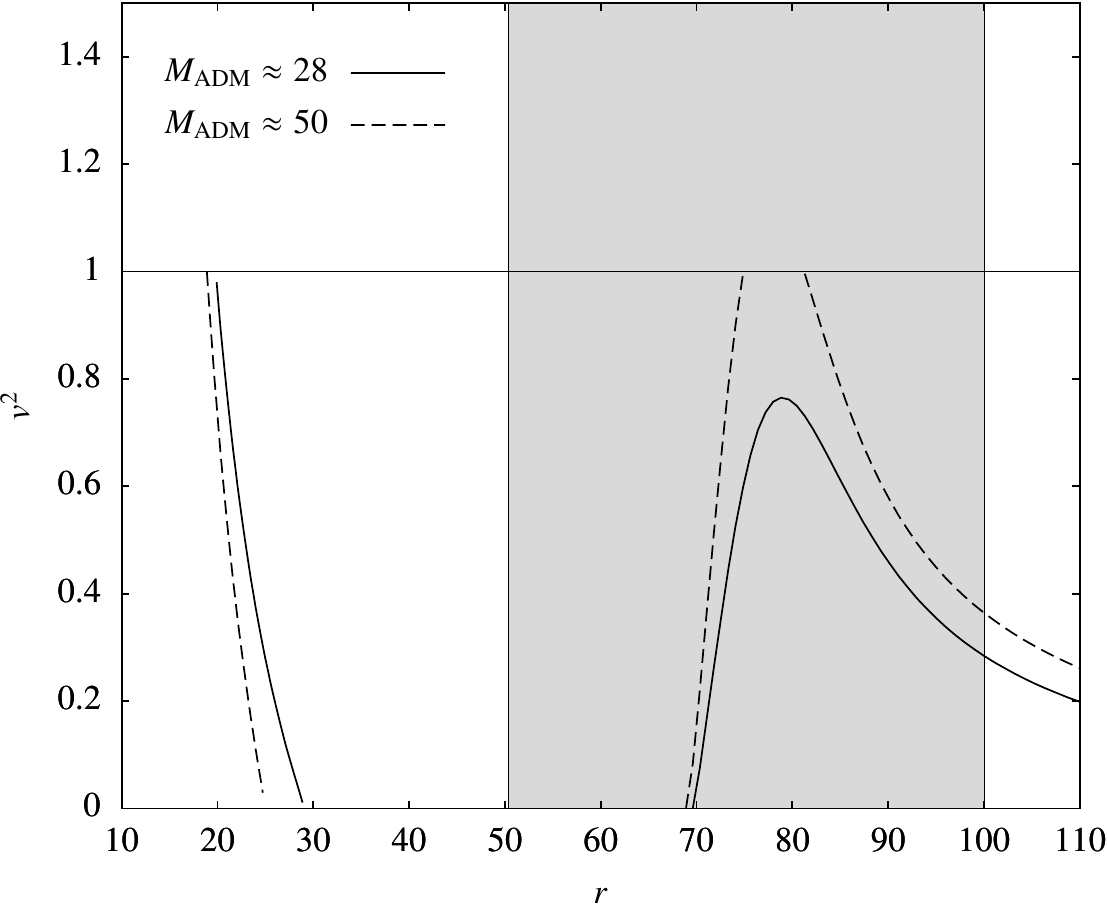}
\caption{\label{photon_orbit}
The square of linear velocity $v^2$ associated with circular geodesics for 2 solutions with the following parameters: $m = 1$, $a = 0$, $\Gamma = 4/3$, $r_1 = 50$, $r_2 = 100$, and different asymptotic masses $M_\mathrm{ADM} \approx 28$ and $M_\mathrm{ADM} \approx 50$. The limit $v^2 = 1$ corresponds to circular photon orbits. The region between $r_1$ and $r_2$ is marked in grey.}
\end{figure}

\begin{figure}
\includegraphics[width=\columnwidth]{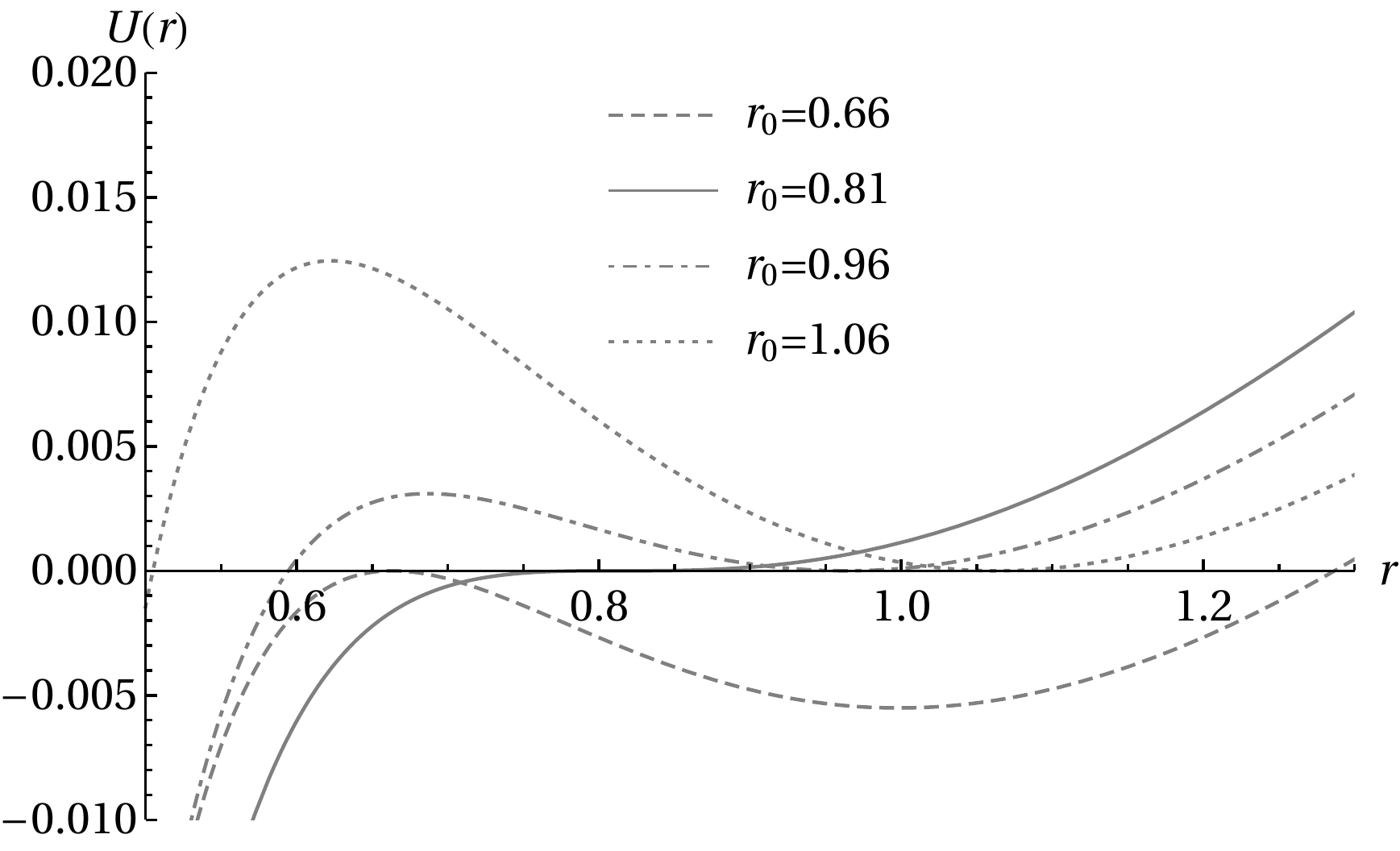}
\caption{\label{iscoT1ErgReg9BH}
The potential $U(r)$ for 4 circular prograde geodesics in the spacetime corresponding to a solution with $m = 1$, $a = 0.96$, $M_\mathrm{ADM} = 1.5$. The ISCO is located at $r_0 \approx 0.81$. The graphs correspond to 4 circular geodesics with radii $r_0 = 0.66, 0.81, 0.96, 1.06$.}
\end{figure}

\begin{figure}
\includegraphics[width=\columnwidth]{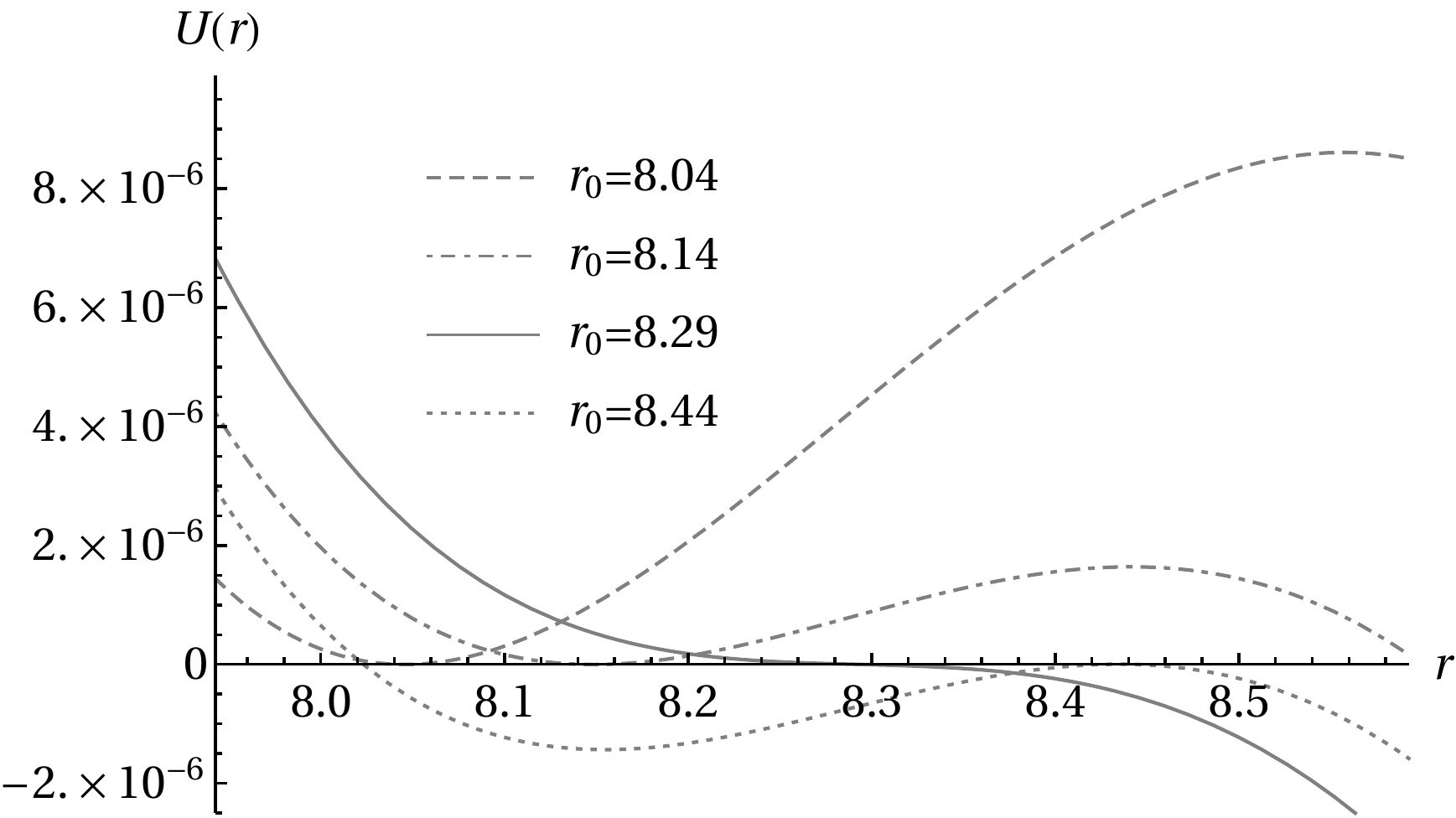}
\caption{\label{iscoT1ErgReg3T1}
The same as in Fig.\ \ref{iscoT1ErgReg9BH}. The circular orbit with $d^2 U(r)/dr^2 = 0$ is located at $r_0 \approx 8.29$. The graphs correspond to 4 circular geodesics with radii $r_0 = 8.04, 8.14, 8.29, 8.44$.}
\end{figure}

\begin{figure}
\includegraphics[width=\columnwidth]{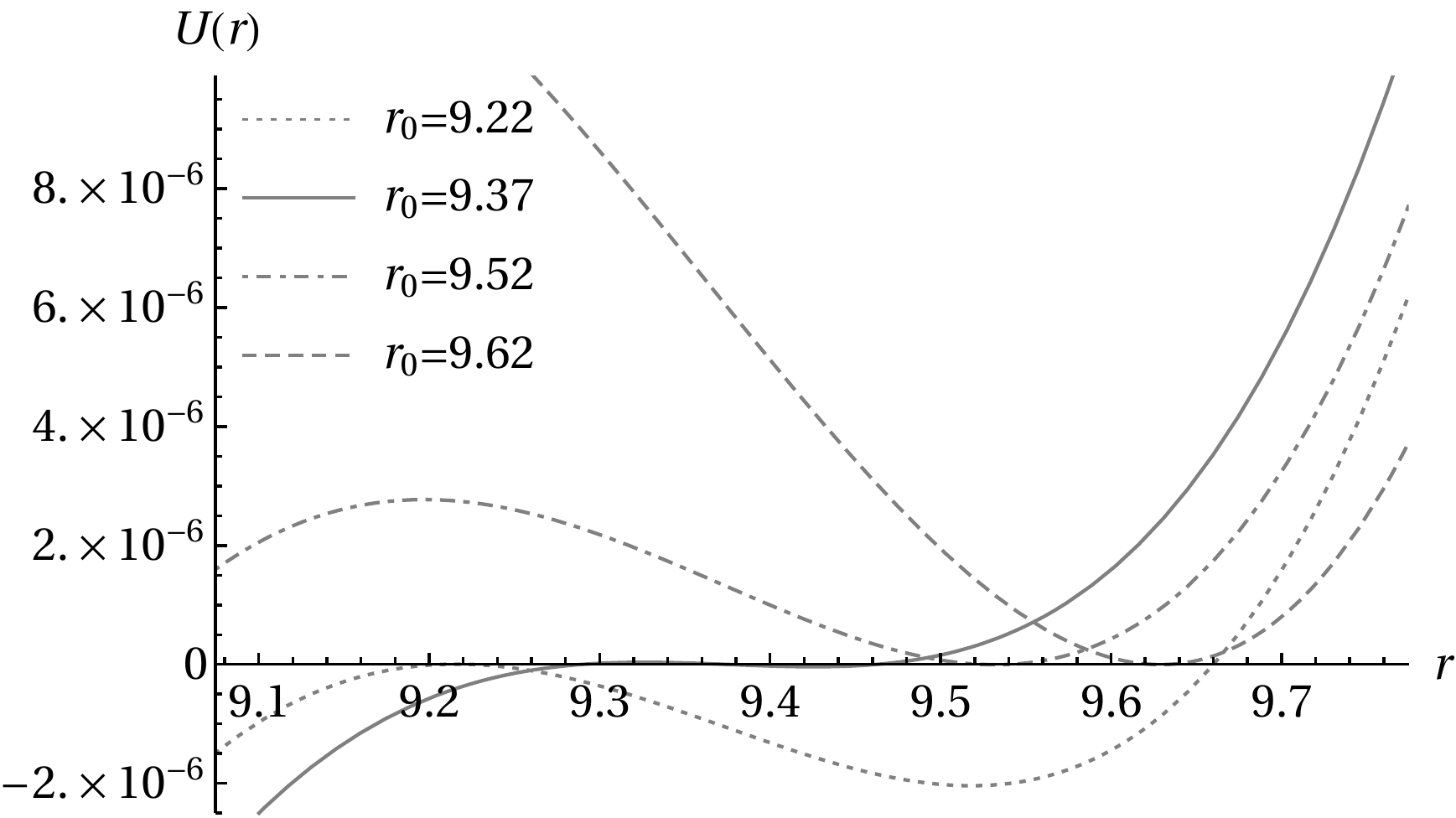}
\caption{\label{iscoT1ErgReg3T2}
The same as in Fig.\ \ref{iscoT1ErgReg9BH}. The circular orbit with $d^2 U(r)/dr^2 = 0$ is located at $r_0 \approx 9.37$. The graphs correspond to 4 circular geodesics with radii $r_0 = 9.22, 9.37, 9.52, 9.62$.}
\end{figure}

\begin{table*}
\caption{\label{tabISCO1} Locations of the Innermost Stable Circular Orbit (ISCO) for a collection of numerical solutions. From left to right the columns report: solution number, the polytropic exponent $\Gamma$, the black-hole spin parameter $a$, the mass of the black hole $M_\mathrm{BH}$, the inner coordinate radius of the torus $r_1$, the inner circumferential radius of the torus $r_\mathrm{c,1}$, the outer coordinate radius of the torus $r_2$, the outer circumferential radius of the torus $r_\mathrm{c,2}$, the coordinate radius of the ISCO $r_\mathrm{ISCO}$, the circumferential radius of the ISCO $r_\mathrm{c,ISCO}$. All solutions were obtained assuming $m = 1$. For solutions I1--I8, the spatial grid resolution around the ISCO is $\Delta r \approx 0.015$; for solutions I9--I12 it is $\Delta r \approx 0.0063$, and in the case of solutions I13--I20 it is $\Delta r \approx 0.022$. For reference: the circumferential radius of the ISCO in the Kerr spacetime with $m = 1$, $a = 0.6$ is $r_\mathrm{c,ISCO} \approx 3.90$; for $m=1$, $a= 0.96$ it is $r_\mathrm{c,ISCO} \approx 2.31$, and for $m=1$, $a= 0.1$ it is $r_\mathrm{c,ISCO} \approx 5.67$.}
\begin{ruledtabular}
\begin{tabular}{c c c c c c c c c c c}
No. & $\Gamma$ & $a$  & $M_\mathrm{BH}$ & $M_\mathrm{ADM}$ & $r_1$ & $r_\mathrm{c,1}$ & $r_2$ & $r_\mathrm{c,2}$ & $r_\mathrm{ISCO}$ & $r_\mathrm{c,ISCO}$ \\
\hline
I1 & $4/3$ & $0.60$ & $1.003$ & $1.050$ & $2.90$ & $4.05$ & $18.10$ & $19.18$ & $2.80$ & $3.95$ \\
I2 & $4/3$ & $0.60$ & $1.005$ & $1.100$ & $2.90$ & $4.07$ & $18.10$ & $19.23$ & $2.81$ & $3.98$ \\
I3  & $4/3$& $0.60$ & $1.011$ & $1.200$ & $2.90$ & $4.12$ & $18.10$ & $19.35$ & $2.86$ & $4.07$ \\
I4 & $4/3$ & $0.60$ & $1.017$ & $1.300$ & $2.90$ & $4.16$ & $18.10$ & $19.46$ & $2.92$ & $4.18$ \\
I5 & $4/3$ & $0.60$ & $1.023$ & $1.400$ & $2.90$ & $4.21$ & $18.10$ & $19.58$ & $3.00$ & $4.31$ \\
I6 & $4/3$ & $0.60$ & $1.029$ & $1.500$ & $2.90$ & $4.26$ & $18.10$ & $19.69$ & $3.09$ & $4.46$ \\
\hline
I7 & $4/3$ & $0.60$ & $1.011$ & $1.200$ & $3.01$ & $4.22$ & $18.10$ & $19.35$ & $2.86$ & $4.07$\\
I8 & $4/3$ & $0.60$ & $1.010$ & $1.200$ & $3.51$ & $4.72$ & $18.10$ & $19.35$ & $2.84$ & $4.05$\\
\hline
I9 & $4/3$ & $0.96$ & $1.001$ & $1.300$ & $6.70$ & $7.97$ & $20.80$ & $22.18$ & $0.82$ & $2.32$\\
I10 & $4/3$ & $0.96$ & $1.001$ & $1.500$ & $6.70$ & $8.10$ & $20.80$ & $22.42$ & $0.81$ & $2.32$\\
I11 & $4/3$ & $0.96$ & $1.003$ & $2.000$ & $6.70$ & $8.43$ & $20.80$ & $23.02$ & $0.81$ & $2.34$\\
I12 & $4/3$ & $0.96$ & $1.005$ & $3.000$ & $6.70$ & $9.14$ & $20.80$ & $24.25$ & $0.80$ & $2.39$\\
\hline
I13 & $4/3$ & $0.10$ & $1.004$ & $1.050$ & $4.81$ & $5.89$ & $20.80$ & $21.87$ & $4.65$ & $5.73$\\
I14 & $4/3$ & $0.10$ & $1.009$ & $1.100$ & $4.81$ & $5.91$ & $20.80$ & $21.93$ & $4.70$ & $5.81$\\
I15 & $4/3$ & $0.10$ & $1.018$ & $1.200$ & $4.81$ & $5.97$ & $20.80$ & $22.04$ & $4.81$ & $5.97$\\
I16 & $4/3$ & $0.10$ & $1.027$ & $1.300$ & $4.81$ & $6.03$ & $20.80$ & $22.16$ & $4.96$ & $6.19$\\
\hline
I17 & $5/3$ & $0.10$ & $1.004$ & $1.050$ & $4.81$ & $5.88$ & $20.80$ & $21.87$ & $4.65$ & $5.73$\\
I18 & $5/3$ & $0.10$ & $1.008$ & $1.100$ & $4.81$ & $5.91$ & $20.80$ & $21.93$ & $4.70$ & $5.80$\\
I19 & $5/3$ & $0.10$ & $1.016$ & $1.200$ & $4.81$ & $5.96$ & $20.80$ & $22.05$ & $4.76$ & $5.91$\\
I20 & $5/3$ & $0.10$ & $1.024$ & $1.300$ & $4.81$ & $6.01$ & $20.80$ & $22.17$ & $4.87$ & $6.08$\\
\end{tabular}
\end{ruledtabular}
\end{table*} 

In this section we study circular geodesics in the equatorial planes of the obtained spacetimes. In particular, we focus on the location of the ISCO, which can be thought of as another quantity characterising black-hole spacetimes. It is usually discussed for the case of the Kerr metric, but it can be also computed for a general axially symmetric metric of the form (\ref{generalmetric}). The relevant formulas can be found for example in \cite{sasaki}. We repeat a few of them here, mainly for completeness. We treat the structure of circular, equatorial geodesics as a probe of the obtained spacetimes, in particular in the context of those geodesic orbits which are contained within the torus (disk).

Consider a circular geodesic at the equatorial plane ($\theta = \pi/2$) in the spacetime endowed with the metric (\ref{generalmetric}). The corresponding angular velocity $\Omega = u^\varphi/u^t$ can be easily found from the geodesic equation. One gets
\begin{equation}
\label{omegageodesic}
\Omega = \frac{-\partial_r g_{t\varphi} \pm \sqrt{(\partial_r g_{t\varphi})^2 -  \partial_r g_{tt} \partial_r g_{\varphi \varphi} }}{\partial_r g_{\varphi \varphi}},
\end{equation}
where the plus and minus signs correspond to prograde and retrograde orbits, respectively.

In order to recall the basics of the stability analysis of circular geodesics, we consider a more general case of a geodesic motion with a potentially non-zero radial component $u^r$. The existence of two Killing vectors $\xi^\mu$ and $\eta^\mu$ implies the existence of two constants of motion: $e = - \xi^\mu u_\mu = - u_t$ and $l = \eta^\mu u_\mu = u_\varphi$ (we adopt the sign convention after \cite{bardeen}). It is a trivial exercise to show that the geodesic equation $u^\mu \nabla_\mu u_\nu = 0$ ensures that $u^\mu \nabla_\mu e = 0$ and $u^\mu \nabla_\mu l = 0$. The normalization of the four-velocity yields
\begin{eqnarray}
 g_{rr} \left( u^r \right)^2 & = & -1 + \frac{e^2}{g_2} \left[ g_{\varphi \varphi} + 2 g_{t \varphi} \frac{l}{e} + g_{tt} \left( \frac{l}{e} \right)^2 \right] \nonumber \\
& \equiv & - U(r),
\label{v}
\end{eqnarray}
where $g_2 = g_{t \varphi}^2 - g_{tt} g_{\varphi \varphi}$. For a circular orbit $u^r = 0$, and hence $U(r) = 0$. The derivative $dU(r)/dr$ computed at a zero of $U(r)$ reads
\begin{eqnarray}
\frac{d U(r)}{dr} & = & \frac{1}{g_2} \left\{ \frac{d g_2}{dr}  - e^2 \left[ \partial_r g_{\varphi \varphi} + 2 \partial_r g_{t \varphi} \frac{l}{e} \right. \right. \nonumber \\
&& \left. \left. + \partial_r g_{tt} \left( \frac{l}{e} \right)^2 \right] \right\}.
\label{vprime}
\end{eqnarray}
The conditions $U(r) = 0$ and $dU(r)/dr = 0$ yield the parameters of circular geodesic orbits. For circular orbits, the constants of motion $e^2$ and $l/e$ can be expressed in terms of $\Omega$ as
\begin{equation}
\label{leratio}
\frac{l}{e} = - \frac{g_{t \varphi} + g_{\varphi \varphi} \Omega}{g_{tt} + g_{t \varphi} \Omega}
\end{equation}
and
\begin{equation}
\label{e2}
e^2 = \frac{(g_{tt} + g_{t\varphi} \Omega)^2}{-g_{tt} - 2 g_{t \varphi} \Omega - g_{\varphi \varphi} \Omega^2}.
\end{equation}
Equations (\ref{leratio}) and (\ref{e2}) follow directly from the normalization condition $u_\mu u^\mu = -1$ and the definitions of $e$ and $l$. Inserting these expressions in Eq.\ (\ref{v}) yields $U(r) = 0$. Analogously, inserting them in Eq.\ (\ref{vprime}) yields a condition for $\Omega$ equivalent to Eq.\ (\ref{omegageodesic}).

In order for a circular orbit to be stable against radial perturbations, it is necessary that $d^2 U(r)/dr^2 > 0$. The second derivative $d^2 U(r)/dr^2$ computed for a circular geodesic, i.e., for $U(r) = 0$, $dU(r)/dr = 0$, reads
\begin{eqnarray}
\frac{d^2 U(r)}{dr^2} & = & \frac{1}{g_2} \left\{ \frac{d^2 g_2}{dr^2}  - e^2 \left[ \partial_{rr} g_{\varphi \varphi} + 2 \partial_{rr} g_{t \varphi} \frac{l}{e}  \right. \right. \nonumber \\
&& \left. \left. + \partial_{rr} g_{tt} \left( \frac{l}{e} \right)^2 \right] \right\},
\label{vbis}
\end{eqnarray}
and of course, in order to check the stability of a given geodesic with a prescribed radius $r$, one has to insert in this formula expressions (\ref{leratio}) and (\ref{e2}), and substitute Eq.\ (\ref{omegageodesic}) for $\Omega$. The condition for the ISCO is obtained as $d^2 U(r)/dr^2 = 0$, and its location in the Kerr spacetime is a standard textbook result.

It is, however, illustrative to observe the behavior of the potential $U(r)$ for different orbits. We plot a couple of examples in Figs.\ \ref{iscokerr} and also \ref{iscoT1ErgReg9BH}, \ref{iscoT1ErgReg3T1}, and \ref{iscoT1ErgReg3T2}.  The graphs in these plots are parametrized by the location of the circular geodesic orbit. Technically, we start by fixing the radius $r_0$ of an orbit, and compute the corresponding angular velocity $\Omega$ from Eq.\ (\ref{omegageodesic}). Then, for $r = r_0$ and the computed value of $\Omega$, we get the constants $e^2$ and $l/e$ from Eqs.\ (\ref{leratio}) and (\ref{e2}). These constant values are then assumed in expression (\ref{v}) for the potential $U(r)$. This means, of course, that $U(r_0) = 0$ and $dU(r_0)/dr = 0$.

Figure \ref{iscokerr} depicts the potentials $U(r)$ computed analytically for four prograde circular orbits with coordinate radii $r_0 = 2.5, 2.7713, 3.1$, and $3.4$ at the equatorial plane in the Kerr spacetime with $m = 1$ and $a = 0.6$. We use quasi-isotropic coordinates. The ISCO is located at $r \approx 2.7713$. As expected, the orbits with $r \gtrapprox 2.7713$ are stable in the sense discussed above.

Figures \ref{iscoa6XrBH}, \ref{iscoa6Xr}, and \ref{iscoa96Xr} depict the quantity 
\begin{equation}
\label{defx}
X(r) = g_2 \frac{d^2 U(r)}{dr^2},
\end{equation}
computed for circular geodesics, as described below Eq.\ (\ref{vbis}). In Figures \ref{iscoa6XrBH} and \ref{iscoa6Xr} we draw $X(r)$ for prograde orbits in the Kerr spacetime with $m = 1$ and $a = 0.6$, i.e., for the case illustrated already in Fig.\ \ref{iscokerr}, but also for a collection of solutions with relatively light tori, characterized by the following parameters: $m = 1$ and $a = 0.6$, $\Gamma = 4/3$, $r_1 = 2.9$, $r_2 = 18.1$, $M_\mathrm{ADM} = 1.05, 1.1, 1.2, 1.3, 1.4$. As expected, for the Kerr solution we have $X(r) = 0$ (ISCO) for $r \approx 2.7713$. Figure \ref{iscoa6Xr} illustrates the deviation of location of the ISCO from the value characteristic for the Kerr spacetime. Similar data obtained for a larger set of solutions are collected in Table \ref{tabISCO1}.

Note that although the coordinate radius of the ISCO $r_\mathrm{ISCO}$ can both grow or decrease with the increasing mass of the torus, the circumferential radius of the ISCO $r_\mathrm{c,ISCO}$ grows with the mass of the torus for all sets of solutions given in Table \ref{tabISCO1}.

If the torus is sufficiently massive, $X(r)$ can have more zeros. There is a zero corresponding to the ISCO (close to the black hole), but there can also be a region with $X(r) < 0$ (unstable geodesics) in the vicinity of the torus. This behavior is illustrated in Fig.\ \ref{iscoa96Xr}. For even more massive tori, one can observe a region (inside the torus), in which the expression $(\partial_r g_{t \varphi})^2 - \partial_r g_{tt} \partial_r g_{\varphi \varphi}$ in Eq.\ (\ref{omegageodesic}) becomes negative, and consequently no circular geodesics exist. A hand-waving explanation of this fact would be to say that in this region the gravity of the torus is larger than that of the black hole, and the net gravitational force cannot act as a centripetal one for the circular geodesic motion. In this case the matter in the torus can still rotate due to the existence of the pressure gradient. Another illustration of this effect can be given by computing the square of the linear velocity for the circular geodesic motion
\begin{equation}
v^2 = \frac{(g_{t\varphi} + g_{\varphi \varphi} \Omega)^2}{g_2},
\end{equation}
where $\Omega$ is given by Eq.\ (\ref{omegageodesic}). This velocity drops to zero at the boundary of the region, where the circular geodesics cease to exist. A graph of $v^2$ corresponding to solutions depicted in Fig.\ \ref{iscoa96Xr} is shown in Fig.\ \ref{V2a96Xr}.

Let us also remark that a necessary condition for a geodesic to be timelike is that $v^2 < 1$. We have, in fact, the standard relation
\begin{equation}
\left( \alpha u^t \right)^2 = \frac{1}{1 - v^2},
\end{equation}
where $\alpha$ denotes the lapse function ($\alpha^2 = g_2/g_{\varphi \varphi}$). An elementary calculation shows that for $v^2 = 1$, one has
\begin{equation}
g_{t t} + 2 g_{t \varphi} \Omega + g_{\varphi \varphi} \Omega^2 = 0,
\end{equation}
and the corresponding geodesic is null. Such circular photon orbits can also be created inside extremely massive disks (tori). An example of two circular photon orbits occurring inside the torus is depicted in Fig.\ \ref{photon_orbit}.

In Figures \ref{iscoT1ErgReg9BH}, \ref{iscoT1ErgReg3T1}, and \ref{iscoT1ErgReg3T2} we show sample graphs of the potential $U(r)$ for a collection of circular geodesics close to three zeros of $X(r)$ for a solution with $m = 1$, $a = 0.96$, $r_1 = 6.7$, $r_2 =20.8 $, $M_\mathrm{ADM} = 1.5$, $\Gamma = 4/3$. The zeros of $X(r)$ are located at $r = 0.81, 8.29, 9.37$. These graphs can be compared with Fig.\ \ref{iscokerr}.

Finally, let us note that the terminology used in this paper is true to the literal meaning of the term ISCO---we refer to the innermost stable circular geodesic orbit as the ISCO, even if there are unstable circular geodesic orbits outside it. One could also define another notion of an innermost stable circular geodesic orbit such that all circular geodesic orbits with larger coordinate radii are stable. In the presence of a sufficiently massive disk (torus) the latter does not have to coincide with the ISCO.

\section{Concluding remarks}
\label{conclusions}

In this paper we investigated several effects of strong gravitational fields in systems consisting of a black hole and a heavy, differentially rotating, perfect fluid torus. Some of these effects are new---for example the parametric bifurcation described in Sec.\ \ref{secbif} or the relativistic effects connected with the volume measures described in Secs.\ \ref{secbif} and \ref{pappus_guldinus}. Others, like the existence of ergoregions associated with massive tori or nonmonotonicity of the circumferential radius at the equatorial plane, have already been reported for rigidly rotating fluids \cite{ansorg:2006, labranche}.

At least some of the efects discussed in this paper seem to be related to each other. For instance, the effect to which we refer as the breaking of the Pappus-Guldinus rule is probably connected with the nonmonotonicity of the circumferential radius, described in Sec.\ \ref{pappus_guldinus}. On the other hand, spotting exact correlations between these effects is difficult.

All abovementioned effects require sufficiently massive tori. In contrast to that, the location of the ISCO, investigated in Sec.\ \ref{circgeodesics}, can be affected by a presence of a moderately massive disk (torus). This fact can, in principle, have astrophysical implications.  

A natural question, not discussed in this paper, is that of the dynamical (nonlinear) stability of presented solutions. The answer to this question can be important both for relatively light and for massive disks (tori). In the former case, it could be astrophysically relevant; in the latter, it could help clarify the physical status of the strong-field effects discussed in this paper.

An analysis similar to the one presented in this article can be also repeated for magnetized tori constructed in \cite{mgfop}. We would expect a lot of similarities, but some effects could turn out to be quantitatively different. One of possible reasons of a slightly different behavior could be due to the fact that strongly magnetized Keplerian self-gravitating tori tend to be much denser in their inner parts, i.e., closer to the black hole.


\begin{acknowledgments}
We would like to thank Andrzej Odrzywo\l{}ek for discussions and improving the performance of the numerical code used to obtain the solutions presented in this paper. We would also like to thank Edward Malec for his careful reading of the manuscript of this paper. This research was carried out with the supercomputer ``Deszno'' purchased thanks to the financial support of the European Regional Development Fund in the framework of the Polish Innovation Economy Operational Program (Contract No.\ POIG.\ 02.01.00-12-023/08). P.\ M.\ was partially supported by the Polish National Science Centre grant No.\ 2017/26/A/ST2/00530. W.\ K.\ acknowledges partial support from the Grant No. 2019-N17/MNS/000026. W.\ D. \ acknowledges partial support from the Grant No.\ 2019-N17/MNS/000018.
\end{acknowledgments}

\end{document}